\documentclass[12pt]{iopart}
\usepackage{iopams}  

\usepackage[dvips]{graphicx}

\def\CC{\mathbb C}

\def\mix{\mathop{\rm mix}}
\def\Ideal{\mathop{\rm Ideal}}

\newtheorem{Thm}{Theorem}

\newtheorem{Lmm}{Lemma}
\newtheorem{Prop}{Proposition}

\begin{document}

\title[Concise and tight finite-key analysis of the BB84 protocol]{Concise and tight security analysis of the Bennett-Brassard 1984 protocol with finite key lengths}

\author{Masahito Hayashi$^{1,2}$ and Toyohiro Tsurumaru$^3$}

\address{
$^1$ Graduate School of Mathematics, Nagoya University, Furocho, Chikusa-ku, Nagoya, 464-860 Japan
}
\address{
$^2$
Centre for Quantum Technologies, National University of Singapore, 3 Science Drive 2, Singapore 117542
}
\address{
$^3$
Mitsubishi Electric Corporation, Information Technology R\&D Center, 5-1-1 Ofuna, Kamakura-shi, Kanagawa, 247-8501, Japan
}

\begin{abstract}
We present a tight security analysis of the Bennett-Brassard 1984 protocol taking into account the finite size effect of key distillation, and achieving unconditional security.
We begin by presenting a concise analysis utilizing the normal approximation of the hypergeometric function.
Then next we show that a similarly tight bound can also be obtained by a rigorous argument without relying on any approximation.

In particular, for the convenience of experimentalists who wish to evaluate the security of their QKD systems,
we also give explicit procedures of our key distillation, and also show how to calculate the secret key rate and the security parameter from a given set of experimental parameters.
Besides the exact values of key rates and security parameters, we also present how to obtain their rough estimates using the normal approximation.
\end{abstract}

\maketitle

\section{Introduction}
The finite size effect is an important issue in practical quantum key distribution (QKD) systems.
The first detailed finite-size analysis for general coherent attacks was given by Hayashi \cite{H07}
using the normal approximation. 
Later, Scarani and Renner \cite{SR08} gave a simple analysis based on the quantum de Finetti Theorem, but their results are valid only against collective attacks.
Matsumoto and Uyematsu also gave a simple analysis \cite{MU10}, but again, essentially valid only for collective attacks.
Later, Tomamichel et al. \cite{TLGR11} gave a tighter bound with unconditional security by using the uncertainty relations (see., e.g., \cite{MU88,RB09}).

In this paper, we present a concise analysis for the Bennett-Brassard 1984 (BB84) protocol \cite{BB84} that takes the finite key effect into account and yields better key generation rates, with and without relying on the normal approximation.
Our analysis is valid for general coherent attacks and thus our results guarantee the unconditional security.
For the sake of simplicity, we consider the case where the sender, Alice, has a perfect single photon source.
We also assume that Alice and the receiver, Bob, calculate an upper bound on the phase error rate of a sifted key, from that of the corresponding sample bits; hence the key generation rate can vary each time Alice and Bob run of the protocol.

Throughout the paper we use the security criteria with universal composability; the same criteria as used by many researcher, particularly by Renner and his coworkers \cite{Renner,R-K}.
Hence our final goal is to show that the trace distance between the actual and the ideal states can be bounded from above.
However, in the mathematical analysis for obtaining upper bounds on the trace distance, we do not use Renner's approach based on the smooth minimum entropy \cite{Renner}.
Instead, we bound the trace distance using the argument by Shor and Preskill \cite{SP00}, as well as its modification by Hayashi \cite{H07}.
In Section \ref{sec:security_criteria}, by using these formalisms, we show that the trace distance can be bounded by using the decoding error probability $P_{\rm ph}$ of the virtual phase error correction;
in other words, the universally composable security can be guaranteed by bounding $P_{\rm ph}$.
To the best of our knowledge, our argument here is the first rigorous treatment of the universally composable security based on the Shor-Preskill formalism,
applicable to linear universal hash functions with variable final key lengths.

As we shall also discuss at the end of Section \ref{sec:security_criteria},
in order to achieve high key generation rates and strong bounds on $P_{\rm ph}$ simultaneously,
it is crucial to estimate the phase error rate $p_{\rm sft}$ of the sifted key with a high accuracy.
Note here that the quantity $p_{\rm sft}$ cannot be measured directly in the BB84 protocol.
Hence in Section \ref{sec:upper_confidence_limit}, we solve an interval estimation problem on $p_{\rm sft}$ using the hypergeometric distribution $P_{\rm hg}$.
Then by using the obtained result, we give explicit bounds on $P_{\rm ph}$ in Section \ref{sec:upper_decoding_error_probability}.
In particular, in order to clarify the argument, we present two versions of analysis:
We first derive a simple bound that we call the {\it straightforward bounds} (Propositions 1 and 2);
and then next give a more complicated bound called the {\it Gaussian bounds} (Theorems 2 and 3), which yield a better final key rate if the raw key is sufficiently large.
For the both types of bounds, we first present a simple analysis based on the normal approximation of the hypergeometric function (Proposition 1 and Theorem 2),
and then next show that a similarly tight bound can also be obtained by a rigorous argument without relying on any approximation (Proposition 2 and Theorem \ref{thm:exact_upperbound}).

Since this paper is not aimed only at theorists, but also at experimentalists who wish to evaluate the security of their QKD systems, we include explicit procedures of security evaluation.
We begin in Section \ref{sec:protocol_description} by explaining explicit procedures of our key distillation.
Then after theoretical arguments of the security, we demonstrate in Section \ref{sec:how_to_use} how to use our theorems to calculate the secret key rate and the security parameter (i.e., an upper bound on the trace distance) from a given set of experimental parameters.
Besides the exact values of key rates and security parameters, we also present how to obtain their rough estimates using the normal approximation.

In order to show that our rates are indeed better than in existing literatures, e.g., Refs. \cite{SR08,TLGR11}, we draw in Section \ref{sec:numerical_results} example curves of key generation rates (Figs. 1 and 2).
There are several reasons for this improvement.
First, our upper bounds are close to the approximated value of the hypergeometric distribution obtained by the normal approximation,
while the existing results \cite{SR08,TLGR11} did not discuss the closeness to the normal approximation.
Second, in our method, the adversary's information is estimated in terms of the Shannon entropy, whereas in \cite{SR08,TLGR11} they use the minimum entropy, which is a lower bound on the Shannon entropy.
Finally, we use an error margin that depends on the measured error rates of sample bits, while in Refs. \cite{SR08,TLGR11} the margin is a constant.

We also treat the sacrifice bit length with the second order coding rate, 
which draws the attention from information theory community \cite{second1,second2,second3}.
The conventional asymptotic theory treats the coding length with the first order coefficient.
It is impossible to treat the approximation value of the best error probability
with the first order coefficient of the coding length.
However, it becomes possible if we consider the coding length up to the second order coefficient.
In this paper, we derive an asymptotic approximation value of the upper bound of 
the universally composable security criterion
when the sacrifice bit length is given as the form $n h(p_{\rm smp}) + \sqrt{n} g(p_{\rm smp})$
with the measured phase error rate,
where a function $g(p_{\rm smp})$ of $p_{\rm smp}$ will be given with a concrete form in Section \ref{sec:upper_confidence_limit} (Theorem \ref{thm:gauss_asymptotic}).

The differences from our previous papers are as follows.
In Refs. \cite{H07}, Hayashi simply approximated the hypergeometric distribution by the normal distribution having the same variance, without showing its validity.
In this paper, we present a rigorous analysis without relying on any approximation (Proposition 2 and Theorem \ref{thm:exact_upperbound}), by using upper bounds on the hypergeometric distribution obtained from the Stirling's formula and inequalities proved in Ref. \cite{Rus,Chvatal}.
As mentioned above, we also included the first rigorous treatment of the universally composable security based on the Shor-Preskill formalism,
applicable to linear universal hash functions with variable final key lengths.

\section{Description of Our QKD Protocol}
\label{sec:protocol_description}
We consider the following type of the BB84 protocol.
This protocol differs from existing versions (e.g., \cite{H07,SR08,MU10}) only in the phase estimation and the privacy amplification steps.

\paragraph{Generation of a Sifted Key and Sample Bits}
Alice and Bob start the protocol with a quantum communication and obtain a sifted key of $n$ bits and sample bits of $l$ bits.
Here we assume that raw key bits are chosen from the uniform distribution.
The sample bits must be selected randomly,
and a sifted key and the sample bits must be measured in different bases.

For example, suppose that Alice and Bob exchange $N$ qubits, choosing the $x$ basis with probability $q$, and the
$z$ basis with $1-q$.
Then, on average, $Nq^2$ bits coincide in the $x$ basis, and $N(1-q)^2$ in the $z$ basis.
By assinging the $x$ basis for a sifted key, and the $z$ basis for sample bits, they have $n=Nq^2$, $l=N(1-q)^2$.\footnote{
In general, however, Alice and Bob may choose bases with different probabilities, and a sifted key and sample bits may be chosen with arbitrary proportions from the two basis.}

\paragraph{Bit Error Correction}
Bob corrects bit errors in his sifted key using a linear error correcting code.
For example, as in Shor-Preskill's case \cite{SP00}, Alice may announce a random bit string XORed with her sifted key;
or alternatively, as in Koashi's case \cite{Koashi}, she may send a syndrome of her sifted key encrypted with a previously shared secret key.
In either case, 
Alice and Bob end up with $n(1-fh(p_{\rm bit}))$ bits of reconciled key $k_{\rm rec}$,
with the bit error rate $p_{\rm bit}$ of a sifted key.
Here $h(x)$ is the binary entropy function defined as $h(x):=-x\log_2x-(1-x)\log_2(1-x)$,
and value $f$ corresponds to the efficiency of the error correcting code used.
For practical codes, $f\simeq1.1$.
It should be noted that here the sizes of bit error correcting codes are independent of the security, and thus Alice and Bob may perform bit error correction by dividing a sifted key $k_{\rm sif}$ of $n$ bits to arbitrarily smaller blocks.

In many cases, one needs to guarantee the correctness of the shared keys, that is, one has to minimize the probability $\epsilon_{\rm cor}$ that Alice's and Bob's secret keys do not match and the protocol does not abort.
One way for minimizing $\epsilon_{\rm cor}$ is that Alice calculates an $r$-bit hash value of her reconciled key $k_{\rm rec}$ using universal$_2$ hash functions.
Then she encrypts it with the one-time pad using a previously shared secret key, and sends it to Bob.
Bob also calculates his own hash value, and if it does not match Alice's, they abort the protocol\footnote{Another possibility is to continue protocol by exchanging supplementary information, such as an additional syndrome, over the public channel, and try bit error correction again.
In such case, the supplementary information also needs to be encrypted with a formerly shared key.}.
By doing this, we have $\epsilon_{\rm cor}\le 2^{-r}$.

\paragraph{Estimation of the number of phase errors in the channel}
In order to use privacy amplification properly and guarantee the security of a secret key, Alice and Bob need to know an upper bound on the number of phase errors occurring in the channel.
It should be noted here that the phase error is a completely different concept from the bit error mentioned above (for details, see Section \ref{sec:security_criteria}).
Since the phase error rate cannot be measured directly in practical QKD systems, 
we estimate its upper bound from the measured error rate of samples.

We denote the number of bit errors occurring in a sample bits by $c$, and the corresponding bit error rate by $p_{\rm smp}(c):=c/l$.
We also call the union of a sifted key and the sample bits {\it total bits}, and 
denote the number of their bit errors by $k$.
Hence the error rate of total bits is given by $p(k):=k/(n+l)$, and that of a sifted key by $p_{\rm sft}(k,c)=(k-c)/n$.
Note here that measuring $c$ corresponds to randomly sampling phase errors in the total bits, because a sifted key and the samples are measured in different bases.
Due to this fact, the measured value of $p_{\rm smp}(c)$ is used to estimate an upper bound on $p_{\rm sft}(k,c)$.
In the asymptotic limit $n,l\to \infty$, Alice and Bob may assume $p_{\rm sft}(k,c)=p_{\rm smp}(c)$.
In practical QKD systems, however, the two values differ in general due to statistical fluctuations.
Thus they obtain a statistically estimated upper bound of $p_{\rm sft}(k,c)$ as a function of the measured value $c$, which we denote by $\hat{p}_{\rm sft}(c)$.
Throughout the paper, we make it a rule to denote an estimated upper bound of a random variable $v$ by $\hat{v}$.
The explicit functional form of $\hat{p}_{{\rm sft},\varepsilon}(c)$ is discussed later,
and is given in Eq. (\ref{eq:def_hat_p_sft}).

\paragraph{Privacy Amplification (PA)}
The estimated phase error rate $\hat{p}_{\rm sft}(c)$ can be used to obtain an upper bound the amount of information that is leaked to Eve.
In order to cancel Eve's information, Alice and Bob perform a classical data processing called privacy amplification on the reconciled key $k_{\rm rec}$ to generate the secret key $k_{\rm sec}$;
very roughly speaking, PA randomizes and shrinks $k_{\rm rec}$ so that Eve's information is canceled by the remaining fraction that is unknown to Eve.
The number of bits to be reduced in this process (sacrifice bits) is determined from $\hat{p}_{\rm sft}(c)$ in the following manner.

We set two limits $c_{\min}$ and $c_{\max}$ ($c_{\min}\le c_{\max})$ on the sample bit error $c$, depending on which Alice and Bob change their procedures.

\begin{itemize}
\item If $c_{\max}<c$, Alice and Bob abort the protocol.
\item If $c_{\min}\le c\le c_{\max}$,
Alice and Bob generate a secret key as the hash value of their sifted key by using a linear and surjective universal$_2$ hash functions.
The number $\alpha(c)$ of sacrifice bits, i.e., the number of bits reduced in PA, is given by
\[
\alpha(c)=n\lceil h\left(\hat{p}_{{\rm sft},\varepsilon}(c+2)\right)\rceil+D.
\]
Here $\lceil x\rceil$ denotes the smallest integer larger than or equal to $x$.
Hence, as a result, they obtain a secret key $k_{\rm sec}$ of $G=n\left[1-fh(p_{\rm bit})\right]-\lceil nh\left(\hat{p}_{{\rm sft},\varepsilon}(c+2)\right)\rceil-D$ bits.
\footnote{Note that key length $G$ of (\ref{eq:def_G_c}) differs from the asymptotic case ($l,n\to \infty$) essentially only in the definition of phase error rate $\hat{p}_{{\rm sft},\varepsilon}(c+2)$.
Hence the estimation of $\hat{p}_{{\rm sft},\varepsilon}(c+2)$ is the key point of our finite size analysis.}

\item If $c<c_{\min}$, Alice and Bob generate a secret key in the same way as above, except that they sacrifice $\alpha(c)=\lceil nh\left(\hat{p}_{{\rm sft},\varepsilon}(c_{\min}+2)\right)\rceil+D$ bits for PA.
As a result, they obtain a secret key $k_{\rm sec}$ of $G=n\left[1-fh(p_{\rm bit})\right]-\lceil nh\left(\hat{p}_{{\rm sft},\varepsilon}(c_{\min}+2)\right)\rceil-D$ bits.
\end{itemize}
Alternatively, we can combine these three case as follows: Define the sacrificed bit length $\alpha(c)$ to be
\begin{equation}
\alpha(c)=\lceil nh\left(\hat{p}_{{\rm sft},\varepsilon}\left(\max[c,c_{\min}]+2\right)\right)\rceil+D.
\label{eq:def_alpha_1}
\end{equation}
If $c\le c_{\max}$, Alice and Bob  sacrifice $\alpha(c)$ bits for PA and obtain the final key of length
\begin{equation}
G(c)=n\left[1-fh(p_{\rm bit})\right]-\alpha(c).
\label{eq:def_G_c}
\end{equation}
If $c\ge c_{\max}$, they abort the protocol.

In practice, the most efficient implementation of PA is to use the Toeplitz matrices:
Alice and Bob select a bit-valued Toeplitz matrix $M$ randomly by communicating over the public channel, multiply it with a reconciled key $k_{\rm rec}$ modulo 2, and obtain the secret key $k_{\rm sec}=Mk_{\rm rec}$ (for details, see., e.g., \cite{Renner,TH10,AT11}).

In this paper, we additionally require the surjectivity for all of hash functions.
To the best of our knowledge, the most efficient implementation of linear and surjective universal$_2$ functions is by using the modified Toeplitz matrix, introduced in \cite{H07,TH10};
in this case we replace $M$ above by a concatenation $M'=(I,T)$ of the (square) identity matrix $I$ and a Toeplitz matrix $T$.
Note that this modification $M'$ is slightly more efficient than $M$ above.
Also note that unlike $M'$, the normal Toeplitz matrix $M$ gives a non-surjective map with a very small but nonzero probability; e.g., for $M$ being an all-zero or all-one matrix.

It should be noted here that, unlike in bit error correction, one is not allowed to perform PA by dividing $k_{\rm rec}$ and $k_{\rm sec}$ into smaller blocks, because doing so will destroy the universal$_2$ property of the (modified) Toeplitz matrix.
Also note here that the both key lengths, $|k_{\rm rec}|=n[1-fh(p_{\rm bit})]$ and $|k_{\rm sec}|=G$, are of order $O(n)$.
If one applies a naive multiplication algorithm, the computational complexity (i.e., the processing time) increases as $O(n^2)$ (i.e., $O(n)$ per key), and thus becomes a severe bottle neck of the key distillation.
This is in fact the most explicit impact of the finite size effect on practical QKD systems.

One way around this problem is to use an efficient multiplication algorithm for a Toeplitz matrix and a vector exploiting the fast Fourier transform (FFT) algorithm (see, e.g., \cite{MatrixTextbook}).
The complexity of this efficient algorithm scales as $O(n\log n)$, or $O(\log n)$ per bit, which can be regarded as a constant in practice.
An actual implementation shows that the throughput exceeds 1Mbps for $|k_{\rm rec}|=10^6$ on software, as demonstrated, e.g., in Ref. \cite{AT11}.

\begin{table}[htbp]
\begin{center}
\begin{tabular}{|c|c|c|c|}
\hline
&total bits & sifted key & sample bits \\ \hline
Number of bits & $n+l$ & $n$ & $l$  \\ \hline
Number of errors & $k$ & $k-c$ & $c$  \\ \hline
Error rate & $p(k)=\frac{k}{n+l}$ & $p_{\rm sft}(k,c)=\frac{k-c}{n}$ 
& $p_{\rm smp}(c)=\frac{c}{l}$  \\ \hline
\begin{tabular}{c}
Estimate of error rate \\
with error probability $\varepsilon$
\end{tabular}
& 
$\hat{p}_{\varepsilon}(c)$ & 
$\hat{p}_{{\rm sft},\varepsilon}(c)$ 
&   \\ \hline
\hline
\end{tabular}
\end{center}
\caption{Notations of the key lengths, total bits, and sample bits.
Functions $\hat{p}_{\varepsilon}(c)$ and $\hat{p}_{{\rm sft},\varepsilon}(c)$ denote the estimated upper bounds of $p(k)$ and $p_{\rm sft}(k,c)$, under the condition that there are $c$ errors in sample bits.
Parameter $\varepsilon$ denotes the probability that the estimation fails. See Section \ref{sec:upper_confidence_limit} for details.
}
\label{table:notations}
\end{table}

\section{Security Criteria of the BB84 Protocol in the finite case}
\label{sec:security_criteria}

\subsection{The security of QKD with universal composability}
We employ the definition of the security of QKD with universal composability in the variable length case \cite{BHLMO}.
In order to guarantee the security for our protocol, 
we need to evaluate the security criteria with universal composability 
after the privacy amplification \cite{R-K}.
In this paper, we apply the above definition with the variable length case
to the final state after the privacy amplification \cite{WMUK}.

For this purpose, we describe all public information by $x$,
including the choice of a hash function (which corresponds, e.g., to ``$f$" of \cite{R-K}),
and the length of the final key (e.g., ``$m$" of \cite{BHLMO}).
However, here we do not restrict ourselves with those two cases; it may contain other public information, e.g., the choice of a code for bit error correction.
Hence the length $m$ of the final key is of course a function of $x$.
We denote the probabilistic distribution of $x$ in the actual protocol by $P_{\rm pub}(x)$.

Then we consider the Hilbert space ${\cal H}_{A}\otimes {\cal H}_{E}\otimes {\cal H}_X$,
consisting of Alice's final key ${\cal H}_A$,
Eve's system ${\cal H}_{E}$, and the public information ${\cal H}_X$.
We define ${\cal H}_A=(\CC^2)^{M}$ with $M$ sufficiently large; so that when $m(x)<M$, Alice uses the (preassigned) subspace of ${\cal H}_A$.
Also, following \cite{Renner}, we define the composite system of $E$ and $X$ to be $E'$, i.e., ${\cal H}_{E'}={\cal H}_{E}\otimes {\cal H}_{X}$.
We denote by $\rho_{A,E|x}$ the state of Alice and Eve after privacy amplification, conditioned on public information $x$.
Hence, the state after privacy amplification takes the form $\rho_{A,E'}=\sum_x P_{\rm pub}(x)\rho_{A,E|x}\otimes |x\rangle\langle x|$.

In this notation, we consider conditional probabilities with respect to length $m$ of the final key.
The actual protocol generates the final key of $m$ bits with probability $P_{\rm len}(m):= \sum_{x:m(x)=m}P_{\rm pub}(x)$.
The public information $x$ obeys the conditional distribution $P(x|m):= \frac{P_{\rm pub}(x)}{P_{\rm len}(m)}$;
hence the conditional actual state given $m$
is a density matrix $\rho_{A,E'|m}:= \sum_{x:m(x)=m} P_{\rm pub}(x|m) \rho_{A,E|x}\otimes |x\rangle \langle x|$.
The corresponding ideal state given $m$ is defined to be $\rho_{\Ideal | m}:=\rho_{A|m}^{\mix} \otimes \rho_{E'|m}$,
where $\rho_{A|m}^{\mix}$ is the completely mixed state in the $m$-qubit subsystem of ${\cal H}_{A}$,
and $\rho_{E'|m}:={\rm Tr}_A \rho_{A,E'|m}$.
Thus, under the condition that the final key length is $m$, the universal compsable security can be guaranteed by bounding the trace distance of these two states, i.e.,
$\left\|\rho_{A,E'|m}-\rho_{\Ideal | m}\right\|_1$ \cite{R-K}.


Parameter $m$ is a random variable in our protocol;
hence following \cite{BHLMO}, we define the universally composable security by bounding the average trace distance $\sum_m P_{\rm len}(m)\left\|\rho_{A,E'|m}-\rho_{\Ideal | m}\right\|_1$.
In this case, it is convenient to define $\rho_{\Ideal}:=\sum_m P_{\rm len}(m)\rho_{\Ideal|m}$.
Then the average trace distance can be rewritten as
\begin{eqnarray}
\left\|\rho_{A,E'}- \rho_{\Ideal}\right\|_1
&=&
\sum_m P_{\rm len}(m) \left\|\rho_{A,E'|m} 
- \rho_{A|m}^{\mix} \otimes \rho_{E'|m}\right\|_1
 \nonumber \\
&=&
\sum_x P_{\rm pub}(x) \left\|\rho_{A,E|x} 
- \rho_{A|m(x)}^{\mix} \otimes \rho_{E|x}\right\|_1
\label{eq:def_uniersal_composability}\\
&\le &
\sum_x P_{\rm pub}(x) \left\|\rho_{A,E|x} - \rho_{A|x} \otimes \rho_{E|x}\right\|_1\label{eq:bound_ideal_real}\\
&&\ \ +
\sum_x P_{\rm pub}(x) \left\|\rho_{A|x} -\rho_{A|m(x)}^{\mix}\right\|_1,
\label{eq:bound_ideal_alices_partial}
\end{eqnarray}
where $\rho_{A|x}:={\rm Tr}_E\rho_{A,E|x}$.
Hence one may instead bound the sum of the second and the third lines.
Here we used the fact that $\rho_{A,E'}=\sum_x P_{\rm pub}(x)\rho_{A,E|x}\otimes|x\rangle\langle x|=\sum_m P_{\rm len}(m)\rho_{A,E'|m}$ for the first equality; and
 $\rho_{E'|m}= \sum_{x:m(x)=m} P_{\rm pub}(x|m) \rho_{E|x}\otimes |x\rangle \langle x|$ for the second equality.
The quantity of (\ref{eq:bound_ideal_alices_partial}) measures the non-uniformity of Alice's final key;
i.e., it gives the averaged distance between Alice's partial state $\rho_{A|x}$ and the ideally mixed state $\rho_{A|m(x)}^{\mix}$.
Note that these two states equal when Alice and Bob choose a surjective hash function, because we assume that Alice's raw key obeys the uniform distribution.
In particular, if Alice and Bob use a hash function family which consists only of surjective functions (such as the modified Toeplitz matirices \cite{H07,TH10} mentioned in the previous section),
it suffices to bound (\ref{eq:bound_ideal_real}) only.


\subsection{Decoding error probability of the virtual phase error correction}
We believe that the above definition of security based on the trace distance is the same as the one used by Renner and others \cite{Renner,R-K}.
Throughout the paper we employ this definition of security.
However, in the remaining part where we actually obtain upper bounds on the trace distance, we do not use Renner's approach based on the smooth minimum entropy \cite{Renner}.
Instead, we bound the trace distance $\|\rho_{A,E|x} - \rho_{A|x} \otimes \rho_{E|x}\|_1$ appearing in (\ref{eq:bound_ideal_real}) using the well-known argument by Shor and Preskill \cite{SP00}, as well as its modification by Hayashi \cite{H07}. 
As we shall see shortly, in these formalisms, the trace distance is bounded from above by using the decoding error probability of the (virtual) phase error correction\footnote{The probability that the (virtual) decoding algorithm fails to give a correct answer.}, which can be identified with the privacy amplification in the actual protocol.
The first step of the proof is to consider a virtual protocol where Alice and Bob correct bit errors as well as phase errors occurring in the quantum channel (under Eve's influence) by using the Calderbank-Shor-Steane (CSS) code.
By correcting these two types of errors, Alice and Bob can guarantee that their virtual channel (obtained as a result of quantum error correction) is noiseless and decoupled from Eve;
thus the key they exchange there is unconditionally secure.
The second step of the proof is to note that, from Eve's view point,
this virtual protocol is completely indistinguishable from the actual protocol.
By using this indistinguishability, the security of the actual protocol follows automatically from that of the virtual protocol.

In these formalisms, phase error correction in the virtual protocol is transformed to a simple classical data processing in the actual protocol.
That is, Alice and Bob do not need to perform phase error correction in the actual protocol; instead it suffices to perform a projection $C_1 \to C_1/C_2$, where $C_1$, $C_2$ are the classical CSS code.
The projection $C_1 \to C_1/C_2$ is often called privacy amplification (PA).
This is why we often identify PA with the virtual phase error correction in this paper\footnote{%
However,
the actual protocol does not necessarily have 
a counterpart for any operation in the virtual protocol.
For example,
the actual protocol has no operation corresponding to 
measurement of the syndrome in the virtual protocol.}.
(In Ref. \cite{TH10}, we have shown that the projection $C_1 \to C_1/C_2$ can be replaced by an $\varepsilon$-almost dual universal$_2$ hash function family.)

The original argument of Shor and Preskill was later improved in Refs. \cite{H06,WMU06},
where it was shown that the virtual phase error correction and the bit error correction can be discussed separately.
In fact the virtual phase error correction is essential for guaranteeing security, while the bit error correction is necessary only for equalizing Alice's and Bob's final keys.
As a result of this observation, 
the trace distance $\|\rho_{A,E|x}-\rho_{A|x} \otimes \rho_{E|x} \|_1$ of  (\ref{eq:bound_ideal_real}) can be bounded as \cite{H07}
\begin{eqnarray}
\left\|\rho_{A,E|x}
-\rho_{A|x} \otimes \rho_{E|x} \right\|_1
\le 2\sqrt2\sqrt
{P_{{\rm ph}|x}},
\label{eq:averaged_distance_concave_condition_x}
\end{eqnarray}
where $P_{{\rm ph}|x}$ denotes the conditional decoding error probability of the virtual phase error correction, given public information $x$.
By taking the average of (\ref{eq:averaged_distance_concave_condition_x}) with respect to $x$, and by noting that the function $a \mapsto \sqrt{a}$ is concave, we have
\begin{eqnarray}
\sum_x P_{\rm pub}(x) 2\sqrt2\sqrt{P_{{\rm ph}|x}}
\le   2\sqrt2\sqrt{\sum_x P(x)_{\rm pub} P_{{\rm ph}|x}}
=   2\sqrt2\sqrt{P_{\rm ph}} ,
\label{eq:averaged_distance_concave}
\end{eqnarray}
where $P_{\rm ph}$ denotes the decoding error probability of the virtual phase error correction.

As to the non-uniformity of the final key given in (\ref{eq:bound_ideal_alices_partial}),
recall that we assumed that Alice's random variable obeys the uniform distribution.
Then the left over hash lemma \cite{HILL,BBCM} yields
\begin{eqnarray}
\sum_x P_{\rm pub}(x)\|\rho_{A|x} -\rho_{A|m(x)}^{\mix}\|_1
\le
\sum_{x} P_{\rm pub}(x) 2^{-\frac{\alpha(x)}{2}} ,
\label{eq:bound_ideal_parital_hash_lemma}
\end{eqnarray}
where $\alpha(x)$ is the number of sacrifice bits in the privacy amplification.

Hence by combining (\ref{eq:def_uniersal_composability})$\sim$(\ref{eq:bound_ideal_alices_partial}), (\ref{eq:averaged_distance_concave}), and (\ref{eq:bound_ideal_parital_hash_lemma})
we obtain
\begin{equation}
\left\|\rho_{A,E'}-\rho_{\Ideal} \right\|_1 
\le  2\sqrt2\sqrt{P_{\rm ph}}
+
\sum_{x} P_{\rm pub}(x) 2^{-\frac{\alpha(x)}{2}}
\label{eq:trace_distance_less_than_Pph_and_exp_alpha}.
\end{equation}
In other words, in order to guarantee the security with universal composability, it suffices to bound the quantity on the right hand side of (\ref{eq:trace_distance_less_than_Pph_and_exp_alpha}).
In particular, as we have noted below (\ref{eq:bound_ideal_alices_partial}), the second term on the right hand side of (\ref{eq:trace_distance_less_than_Pph_and_exp_alpha}) is exactly zero when all of the hash functions are surjective; 
in this case the above inequality is replaced by 
\begin{eqnarray}
\|\rho_{A,E'}-\rho_{\Ideal} \|_1 
\le 2\sqrt2\sqrt{ P_{\rm ph}}
\label{eq:trace_distance_less_than_Pph}.
\end{eqnarray}
Hence, in order to guarantee the universally composable security, it suffices to bound $P_{\rm ph}$.

\subsection{Conditional decoding error probability given $k$}
\label{sec:upper_bound_Pph_each_k}
In this subsection we show that, in order to bound the decoding error probability $P_{\rm ph}$ of the virtual phase error correction, it is sufficient to bound $P_{{\rm ph}|k}$ for all $k$,
where $P_{{\rm ph}|k}$ denotes the corresponding conditional probability given $k$.
We also show that a bound on $P_{{\rm ph}|k}$ can be given in a concise form using the hypergeometric distribution $P_{\rm hg}(c|k)$ and binary entropies.

First note that, without loss of generality, Eve's eavesdropping strategy can be described by the probability distribution $Q_{\rm Eve}(k)$ of $k$, which is the number of errors in the total bits $n+l$\footnote{%
In the general setting, Eve is allowed to use the superposition among different intgers $k$.
In order to treat such a case, we introduce 
the distribution $Q_{\rm Eve}(k)$ here.}.
Then $P_{\rm ph}$ can be rewritten as $P_{\rm ph}= \sum_k Q_{\rm Eve}(k) P_{{\rm ph}|k}$, where $P_{{\rm ph}|k}$ denotes the conditional decoding error probability given $k$.

Next we consider the conditional probability $P_{\rm hg}(c|k)$ of $c$ given $k$; i.e., the probability that $c$ bits of errors are found in sample bits when there are $k$ errors in the total bits.
Since sample bits are sampled without replacement, $c$ obeys the hypergeometric distribution for a fixed value of $k$:
\begin{equation}
P_{\rm hg}(c|k):=\frac{{n\choose k-c}{l\choose c}}{{n+l\choose k}},
\end{equation}
with the average $\bar{c}$ and the deviation $\sigma$ given by
\begin{equation}
\bar{c}(k):=\frac{lk}{n+l},\quad
\sigma_{n,l}(k)^2:=\frac{knl(n+l-k)}{(n+l)^2(n+l-1)}.
\label{eq:def_barc_sigma}
\end{equation}
In the following, 
$\sigma_{n,l}(k)^2$ is simplified to $\sigma(k)^2$.
Hence values of $k,c$ occurs with probability $Q_{\rm Eve}(k)P_{\rm hg}(c|k)$.
(Here sample bits are sampled without replacement simply because one cannot measure both the phase and the bit values of a qubit simultaneously, and thus Alice and Bob cannot reuse the sample bits as a sifted key.
If one could somehow sample them with replacements, the hypergeometric distribution here would of course be replaced by the binomial distribution, which is much simpler.)

Finally we consider the conditional decoding error probability $P_{{\rm ph}|k,c}$ for fixed values of $k$ and $c$.
In this case, the number of phase error patterns of total bits is bounded from above by $2^{nh\left((k-c)/n\right)}$ (see, e.g., Lemma 4.2.2, Ref. \cite{Justesen}).
Due to the construction of the procotocl, the number of the sacrificed bits $\alpha(c)$ is fixed.
As we have shown in Ref. \cite{TH10}, if Alice and Bob use a linear universal$_2$ hash function family for PA in the actual protocol, 
it can be considered as the situation in the virtual protocol where they use a 2-almost universal$_2$ linear code family for phase error correction
(i.e., a linear 2-almost universal$_2$ hash function family is used as the syndrome function for correcting phase errors).
Then the decoding error probability $P_{{\rm ph}|k,c}$ of the virtual phase error correction can be bounded as
\begin{eqnarray}
P_{{\rm ph}|k,c}&\le& S_{\rm pa}(k,c) := 2\cdot2^{\left[g(k,c)\right]^-}=2^{\left[g(k,c)\right]^-+1},
\label{eq:upper_bound_S_pa}\\
g(k,c):&=&nh\left((k-c)/n\right)-\alpha(c)\nonumber\\
&=&nh\left((k-c)/n\right)-nh\left(\hat{p}_{\rm sft}(c+2)\right)-D
\label{eq:def_g_k_c}
\\
&=&nh\left(p_{\rm sfc}(k,c)\right)-nh\left(\hat{p}_{\rm sft}(c+2)\right)-D,\nonumber
\end{eqnarray}
where $[x]^-:=\min(x,0)$.
It is easy to see that Inequality (\ref{eq:upper_bound_S_pa}) holds when the completely random matrices (a type of universal$_2$ hash functions) are used for PA, as in Koashi's case \cite{Koashi}.
It is also shown to hold when the Toeplitz matrices (another universal$_2$ hash function family) are used for PA, by using the fact that dual matrices of the Toeplitz matrices generate universal$_2$ hash functions \cite{H07}.
More generally, in Ref. \cite{TH10},  we have further shown that Inequality (\ref{eq:upper_bound_S_pa}) is valid when an arbitrary family of universal$_2$ functions is used for PA.

Hence, to summarize, under Eve's strategy $Q_{\rm Eve}(k)$, error numbers $k,c$ are distributed by $Q_{\rm Eve}(k)P_{\rm hg}(c|k)$.
For fixed values of $k,c$, the virtual phase error correction fails with a probability less than $S_{\rm pa}(k,c)$ given in (\ref{eq:upper_bound_S_pa}).
Combining these probabilities, we see that the decoding error probability $P_{\rm ph}$ of the virtual phase correction can be bounded as
\begin{eqnarray}
P_{\rm ph}=\sum_k Q_{\rm Eve}(k)P_{\rm ph|k} 
&\le& \sum_k \sum_c Q_{\rm Eve}(k) P_{\rm hg}(c|k)S_{\rm pa}(k,c) \\
&=& \sum_k Q_{\rm Eve}(k) S_{\rm av}(k)
\le \max_k S_{\rm av}(k),
\label{eq:bound_Pph_by_max_Sk}
\end{eqnarray}
where
$S_{\rm av}(k)$ is defined by
\begin{equation}
S_{\rm av}(k):=\sum_{c=0}^{c_{\max}}
P_{\rm hg}(c|k)S_{\rm pa}(k,c).
\label{eq:S_av_bound_1}
\end{equation}
Since Eve's strategy $Q_{\rm Eve}(k)$ can be arbitrary,
$P_{\rm ph}$ can be bounded if and only if $\max_k S_{\rm av}(k)$ is bounded.
Hence in what follows, we will concentrate on obtaining upper bounds on $\max_k S_{\rm av}(k)$.

As one can see from the definition of $S_{\rm pa}(k,c)$ in (\ref{eq:upper_bound_S_pa}), (\ref{eq:def_g_k_c}), a straightforward way of minimizing $\max_k S_{\rm av}(k)$ is to define the function $\hat{p}_{\rm sft}(c)$ so that it always gives a large value;
this corresponds to the situation where, looking at $c$, Alice and Bob always give a pessimistic estimate $\hat{p}_{\rm sft}(c)$ that is much larger than the actual value $p_{\rm sft}(k,c)$.
However, as one can see from the definition of  $\alpha(c)$ in (\ref{eq:def_alpha_1}) and the final key length $G$ given in the previous section, a large $\hat{p}_{\rm sft}(c)$ results in a poor key generation rate.
Rather, in order to achieve high key generation rates and the high-level security simultaneously,
one needs to minimize $\max_k S_{\rm av}(k)$ by considering the contributions of the two factors, $P_{\rm hg}(k|c)$ and $S_{\rm pa}(k,c)$.
Hence we define $\hat{p}_{\rm sft}(c)$ so that it becomes as close as possible (and larger) to the actual value $p_{\rm sft}(k,c)$, in the regions of $k,c$ where $P_{\rm hg}(c|k)$ is not negligible.
This is equivalent to the estimation problem of an upper bound of $p_{\rm sft}(k,c)$:
\begin{enumerate}
\item For a given $c$, we give a suitable choice of the estimated value $\hat{p}_{\rm sft}(c)$ for the phase error rate of a sifted key.
Alice and Bob use this value to calculate the value of $\alpha(c)$ of (\ref{eq:def_alpha_1}), and obtain the final key length $G$.
This will be done in Section \ref{sec:upper_confidence_limit}.
\item With the suitable choice of $\hat{p}_{\rm sft}(c)$, we obtain a universal upper bound on the RHS of (\ref{eq:S_av_bound_1}) that is independent of $k$, and thus an upper bound of $P_{\rm ph}$\footnote{A similar analysis was given by Fung et al. \cite{Fung10}. However, they seem to evaluate $P_{\rm hg}(c|k)S_{\rm pa}(k,c)$ without the summation. This corresponds to the probability that a certain set of values $k$ and $c$ occur {\it and then} the virtual phase error correction by Alice and Bob fails.}.
This will be done in Section \ref{sec:upper_decoding_error_probability}.
\end{enumerate}


\section{Upper confidence limit on the phase error rate $p_{\rm sft}(k,c)$}
\label{sec:upper_confidence_limit}

Now let us turn to the definition of $\hat{p}_{\rm sft}(c)$.
As mentioned above, since length $l$ of sample bits is finite in practical QKD systems, the phase error rate of a sifted key $p_{\rm sft}(k,c)$ deviates from that of sample bits, $p_{\rm smp}(c)$, due to statistical fluctuations.
Hence, in order to guarantee the security by privacy amplification, instead of $p_{\rm smp}(c)$, one needs to use the estimated upper bound $\hat{p}_{\rm sft}(c)$ of $p_{\rm sft}(k,c)$, defined with the statistical effect taken into account. 

As long as $p_{\rm sft}(k,c)$ is estimated larger than the actual value, i.e., $\hat{p}_{\rm sft}(c)>p_{\rm sft}(k,c)$,
there is no loss of security, because then, more information is erased by the privacy amplification than is actually leaked to Eve.
On the other hand, however, one needs to avoid a situation where $p_{\rm sft}(k,c)$ is estimated smaller as $\hat{p}_{\rm sft}(c)\le p_{\rm sft}(k,c)$. In such a case, the privacy amplification of the previous section does not work since $[g(k,c)]^-=0$.
Hence, at least as a necessary condition, the function $\hat{p}_{\rm sft}$ needs to satisfy that
\begin{equation}
{\rm Pr}_k\left\{\ c\ \left|\ \hat{p}_{\rm sft}(c) \ge p_{\rm sft}(k,c)\ \right.\right\}
> 1-\varepsilon\ \ {\rm for}\ \forall k,
\label{eq:estimation_necessary_condition}
\end{equation}
where ${\rm Pr}_k\left\{c|Q\right\}$ denotes the probability that $c$ occurs satisfying a condition $Q$, under the hypergeometric distribution $P_{\rm hg}(c|k)$.
In order to maximize the key generation rate for fixed values of $l,n$, we wish to minimize $\hat{p}_{\rm sft}(c)$ as small as possible.
In statistics, this corresponds to an interval estimation problem.
That is, finding $\hat{p}_{\rm sft}(c)$ satisfying (\ref{eq:estimation_necessary_condition}) is to obtain an upper confidence limit on  $p_{\rm sft}(k,c)$ from an observed value of $c$, with significance level $\varepsilon$ (see, e.g., \cite{Hoel}).

In the following,
we derive the minimum estimate 
$\hat{p}_{{\rm sft},\varepsilon}(c)=\hat{p}_{{\rm sft}}(c)$ satisfying the condition 
(\ref{eq:estimation_necessary_condition})
under the normal approximation of $P_{\rm hg}(c|k)$
by employing interval estimation of $k$.
Although there is a standard procedure found in every textbook for this analysis (e.g., \cite{Hoel}), we reproduce it below for the sake of explanation.
First we define 
the normal distribution function by
\begin{equation}
\Phi(x):=\frac1{\sqrt{2\pi}}\int_x^\infty\exp(-y^2/2)dy,
\label{eq:def_normal_distribution_function}
\end{equation}
and $s(\varepsilon)$ as the deviation corresponding to $\varepsilon$, e.g.,
\begin{equation}
s(\varepsilon)=\Phi^{-1}(\varepsilon)
\label{eq:def_s_epsilon}
\end{equation}
such that $\varepsilon=\Phi(s(\varepsilon))$.
In what follows, we often abbreviate $s(\varepsilon)$ to $s$.
Then, by applying the normal approximation to $P_{\rm hg}(c|k)$, we have the relation
\begin{equation}
{\rm Pr}_k\left\{\ c\ \left|\ c \ge \bar{c}(k)-s(\varepsilon)\sigma(k)\ \right.\right\}>1- \varepsilon
\label{eq:condition_Pr_k_approx}
\end{equation}
for any integer $k$;
that is,
$c \ge \bar{c}(k)-s(\varepsilon)\sigma(k)$ holds 
at least with probability $1-\varepsilon$ for any integer $k$.
Note that this condition is equivalent to
$(c- \bar{c}(k))^2 \le s(\varepsilon)^2\sigma(k)^2$
or $c\ge \bar{c}(k)$.
We rewrite this condition further as
\begin{equation}
\left(p_{\rm smp}-p\right)^2 \le 4\gamma p(1-p),
\hbox{ or }
p_{\rm smp} \ge p
\label{eq:p_hat_quadratic_eq}
\end{equation}
where $p=k/(n+l)$, $p_{\rm smp}(c)=c/l$,
and
\begin{equation}
\gamma:=\frac{s(\varepsilon)^2n}{4l(n+l-1)}.
\label{eq:def_gamma}
\end{equation}

The condition (\ref{eq:p_hat_quadratic_eq}) is equivalent to
$p \le \hat{p}_{\varepsilon}(c) $,
where $\hat{p}_{\varepsilon}(c)$ is a solution of $\left(p_{\rm smp}-\hat{p}_{\varepsilon}\right)^2 = 4\gamma \hat{p}_{\varepsilon}(1-\hat{p}_{\varepsilon})$ given by
\begin{equation}
\hat{p}_{\varepsilon}(c) := \frac1{1+4\gamma}
\left(
p_{\rm smp}
+2\gamma
+2\sqrt{\gamma\left\{p_{\rm smp}\left(1-p_{\rm smp}\right)+\gamma\right\}}
\right).
\label{eq:def_hat_p}
\end{equation}
That is,
$k/(n+l)=p \le \hat{p}_{\varepsilon}(c) $ holds 
at least with probability $1-\varepsilon$ for any integer $k$.
In other words, the rate 
$\hat{p}_{\varepsilon}(c) $
gives the upper bound of one-sided interval estimation of $p=k/(n+l)$.
Using this estimate, we define another function 
\begin{equation}
\hat{p}_{{\rm sft},\varepsilon}(c):=
(\hat{p}_{\varepsilon}(c) (n+l)-c)/n
=
\frac{(n+l)\hat{p}_\varepsilon(c)-lp_{\rm smp}(c)}n.
\label{eq:def_hat_p_sft}
\end{equation}
Then, again, the inequality
$\hat{p}_{{\rm sft},\varepsilon}(c) \ge p_{\rm sft}(k,c)=(k-c)/n$
holds at least with probability $1-\varepsilon$ for any integer $k$.
As a result, by choosing $\hat{p}_{\rm sft}(c)$ as $\hat{p}_{{\rm sft},\varepsilon}(c)$, we can satisfy the condition (\ref{eq:estimation_necessary_condition}).
Throughout the paper, we will use these definitions of $\hat{p}_{\varepsilon}(c)$ and $\hat{p}_{{\rm sft},\varepsilon}(c)$ in calculating $\alpha(c)$.

Now two remarks are in order.
First, if there are sufficiently many samples (i.e., with $l$ large and thus $\gamma$ sufficiently small), the error number $c$ has roughly the same distribution, irrespective of whether the samples are picked up with or without replacement.
In such a case, as we mentioned under Eq. (\ref{eq:def_barc_sigma}), the hypergeometric distribution $P_{\rm hg}(c|k)$ can be approximated by the binomial distribution.
Indeed, to the first order of $\sqrt{\gamma}$, the estimated value $\hat{p}_{\varepsilon}(c)$  of Eq. (\ref{eq:def_hat_p}) can be approximated as
\begin{eqnarray*}
\hat{p}_{\varepsilon}(c)&\simeq& p_{\rm smp}(c)+\frac{s}{l}\sqrt{\frac{n}{n+l-1}}\,\sigma_{\rm bin}(c)\\
&=& p_{\rm smp}(c)+\frac{s}{l}\sqrt{\frac{n}{n+l-1}}\sqrt{lp_{\rm smp}(c)(1-p_{\rm smp}(c))},
\end{eqnarray*}
where $\sigma_{\rm bin}(c):=\sqrt{lp_{\rm smp}(c)(1-p_{\rm smp}(c))}$ denotes the deviation of the binomial distribution with the error rate of the sample bits being $p_{\rm smp}(c)=c/l$.
Furthermore, by using the inequality $
p_{\rm smp}(c)+\frac{s}{l}\sqrt{\frac{n}{n+l-1}}\,\sigma_{\rm bin}(c)
\le p_{\rm smp}(c)+\frac{s}{l}\sigma_{\rm bin}(c)$, and by noting that the larger $\hat{p}_{\varepsilon}(c)$ always gives better a security bound,
we can instead use a simpler approximation given by
\begin{equation}
\hat{p}_{\varepsilon}(c)\simeq p_{\rm smp}(c)+\frac{s}{l}\sigma_{\rm bin}(c),
\label{eq:approx_hat_p}
\end{equation}
The approximated upper bound of (\ref{eq:approx_hat_p}) can also be obtained by an argument similar to the above, with the hypergeometric distribution replaced by the binomial distribution.
This means that, for $l$ sufficiently large, one can conclude that the phase error rate $p(k,c)$ of the total bits can be bounded from above by $\hat{p}_{\varepsilon}(c)$ of (\ref{eq:approx_hat_p}), which is simply the measured error rate $p_{\rm smp}(c)$ of the samples, plus $s$ times its standard deviation $\frac{s}{l}\sigma_{\rm bin}$.
The actual value deviates this bound only with a probability less than $\Phi(s)$; or in other words, this estimation fails only with a probability less than $\Phi(s)$.

\section{Upper bounds on the decoding error probability $ P_{\rm ph}$}
\label{sec:upper_decoding_error_probability}

Throughout the paper, 
we assume that Alice and Bob perform the protocol specified in Section \ref{sec:protocol_description},
using the estimated upper bound $\hat{p}_{{\rm sft},\varepsilon}(c)$ of (\ref{eq:def_hat_p}) and (\ref{eq:def_hat_p_sft}), obtained in the previous section.
That is, we here substitute $\hat{p}_{{\rm sft},\varepsilon}(c)$ for $\hat{p}_{{\rm sft}}(c)$ in (\ref{eq:def_alpha_1}),
and as a result of that, Alice and Bob use sacrifice bits of $\alpha(c)=h\left(\hat{p}_{{\rm sft},\varepsilon}(\max[c,c_{\min}])\right)+D$ in the PA step.
In this setting, we evaluate the decoding error probability evaluate $P_{\rm ph}$ and obtain several upper bounds.

\subsection{The Straightforward Upper Bounds}
In Section \ref{sec:upper_bound_Pph_each_k}, we showed that, in order to bound $P_{\rm ph}$, it suffices to bound $S_{\rm av}(k)$ of (\ref{eq:S_av_bound_1}) for all values of $k$.
In this subsection, we first present a simple evaluation of $P_{\rm ph}$, where we divide the summation $S_{\rm av}(k)$, given in (\ref{eq:S_av_bound_1}), into two regions of $c$.
This method is similar to those used in preceding literature \cite{SR08,MU10}, and we call it here the {\it straightforward method}.

For each value of $k$, we set the boundary value $c_{\rm bnd}(k):=\lfloor\bar{c}(k)-s\sigma(k)\rfloor$, and divide the summation of (\ref{eq:S_av_bound_1}) as
\begin{eqnarray}
S_{\rm av}(k)&=&\sum_{c=0}^{c_{\max}} P_{\rm hg}(c|k)S_{\rm pa}(k,c)\\
&\le&\sum_{c=0}^{\lfloor \bar{c}(k)-s\sigma(k) \rfloor} P_{\rm hg}(c|k)+\sum_{c=\lfloor\bar{c}(k)-s\sigma(k)\rfloor+1}^{c_{\max}}P_{\rm hg}(c|k)S_{\rm pa}(k,c)
\label{eq:straightforward_division_two_regions2}
\\
&\le&\sum_{c=0}^{\lfloor\bar{c}(k)-s\sigma(k) \rfloor} P_{\rm hg}(c|k)+\max_{c\in[\bar{c}(k)-s\sigma(k),c_{\max}]}S_{\rm pa}(k,c).
\label{eq:straightforward_division_two_regions}
\end{eqnarray}
(In what follows, we often write  $\bar{c}$, $\sigma$, $s$ instead of $\bar{c}(k)$, $\sigma(k)$, $s(\varepsilon)$.)
Then, by using the properties of $\hat{p}_{{\rm sft},\varepsilon}(c)$ given in the preceding section, 
the two terms of (\ref{eq:straightforward_division_two_regions}) can be evaluated as follows:
\begin{enumerate}
\item
The first summation of (\ref{eq:straightforward_division_two_regions}) is the probability ${\rm Pr}_k\left\{\ c\ \left|\ c < \bar{c}(k)-s(\varepsilon)\sigma(k)\ \right.\right\}$.
As we have shown in the preceding section, this term is less than $\varepsilon$  (see (\ref{eq:condition_Pr_k_approx})),
if one applies the normal approximation to $P_{\rm hg}(c|k)$.
To put it more explicitly, apply the normal approximation of the form:
\begin{equation}
\sum_{c=a}^bP_{\rm hg}(c|k)\simeq\frac1{\sqrt{2\pi}}\int_{\zeta_a}^{\zeta_b}e^{-x/2}dx
\label{eq:def_normal_approx_hypergeo}
\end{equation}
with $\zeta_c:=(c-\bar{c}(k))/\sigma(k)$.
Then it follows that the first term of  (\ref{eq:straightforward_division_two_regions}) is less than $\Phi(s(\varepsilon))=\varepsilon$,
where $\Phi(s)$ is the normal distribution function given in (\ref{eq:def_normal_distribution_function}).
\item
In the second term of (\ref{eq:straightforward_division_two_regions}), the function $S_{\rm pa}(k,c)=2^{\left[g(k,c)\right]^-+1}$ is maximized at $c=\bar{c}(k)-s\sigma(k)$, because $g(k,c)$, defined in (\ref{eq:def_g_k_c}), is decreasing with $c$.
Also note that 
\[
\hat{p}_{{\rm sft},\varepsilon}(\bar{c}(k)-s\sigma(k))=p_{\rm sft}(k,\bar{c}(k)-s\sigma(k))
\]
holds by the definition of $\hat{p}_{{\rm sft},\varepsilon}(c)$, given in (\ref{eq:def_hat_p}) and (\ref{eq:def_hat_p_sft}).%
\footnote{In fact, this is exactly the way we planned when we defined $\hat{p}_{{\rm sft},\varepsilon}(c)$:
As mentioned in sentences below (\ref{eq:P_ph_upper_bound_by_delta}),
the function $\hat{p}_{\varepsilon}(c)$ is defined so that the condition
$\hat{p}_{\varepsilon}(\bar{c}(k)-s\sigma(k))=p(k)$
is satisfied for all $k$.
This condition is equivalent to $\hat{p}_{{\rm sft},\varepsilon}(\bar{c}(k)-s\sigma(k))=p_{\rm sft}(k,\bar{c}(k)-s\sigma(k))$, due to definitions of $\hat{p}_{{\rm sft},\varepsilon}(c)$ and $p_{\rm sft}(k,c)$ given in (\ref{eq:def_hat_p_sft}) and in Table \ref{table:notations}.}
Thus from (\ref{eq:def_g_k_c}), we have
\begin{eqnarray*}
\lefteqn{g\left(k,\bar{c}(k)-s\sigma(k)\right)=nh\left(p_{\rm sft}(k,\bar{c}(k)-s\sigma(k))\right)-\alpha\left(\bar{c}(k)-s\sigma(k)\right)}\\
&\le&nh\left(p_{\rm sft}(k,\bar{c}(k)-s\sigma(k))\right)-nh\left(\hat{p}_{{\rm sft},\varepsilon}\left(\bar{c}(k)-s\sigma(k)\right)\right)-D=-D
\end{eqnarray*}
For the inequality of the second line, we used the fact that $\alpha(c)=h\left(\hat{p}_{{\rm sft},\varepsilon}(\max[c,c_{\min}]+2)\right)\ge h\left(\hat{p}_{{\rm sft},\varepsilon}(c)\right)$.
This means that the second summation of (\ref{eq:straightforward_division_two_regions}) can be bounded by $2^{-D+1}$.
We remark that, unlike the first term of (\ref{eq:straightforward_division_two_regions}), this upper bound is valid without relying on the normal approximation.
\end{enumerate}
Note here that the both bounds are valid for all values of $k$.
Hence by combining these two upper bounds,
%
we obtain the following proposition.
\begin{Prop}\label{thm:st_upperbound}
For a given $\varepsilon$ (and the corresponding $s(\varepsilon)=\Phi^{-1}(\varepsilon)$), 
suppose that $c_{\min}\le c_{\max}$, and that Alice and Bob perform the QKD protocol specified in Section \ref{sec:protocol_description}.
%
%
Then by applying the normal approximation to $P_{\rm hg}(c|k)$, 
$P_{\rm ph}$ can be bounded as
\begin{equation}
P_{\rm ph}
\le \max_k S_{\rm av}(k)
\le
\varepsilon +2^{-D+1}.
\label{eq:P_ph_upper_bound_by_Delta}
\end{equation}
\end{Prop}

\setcounter{footnote}{0}

If one wishes to bound $P_{\rm ph}$ by a certain value, say $P_{\max}$,
a convenient choice of parameters is $\varepsilon=2^{-D+1}=\frac12P_{\max}$,
or equivalently, $D=2-\log_2P_{\max}$ and $s=\Phi^{-1}(\varepsilon)=\Phi^{-1}\left(\frac12P_{\max}\right)$.%
\footnote{Of course, the optimal choice is to let $\varepsilon=aP_{\max}$ and $2^{-D+1}=(1-a)P_{\max}$, and then find the optimal $0<a<1$ that yields the largest key generation rate.
However, we do not pursue this optimality in the rest of the paper, since varying $a$ contributes very little to the key rate in typical situations.}
Then Inequality (\ref{eq:trace_distance_less_than_Pph}) guarantees that the trace distance is bounded as $\|\rho_{A,E'}-\rho_{\Ideal} \|_1\le 2\sqrt2\sqrt{P_{\max}}$,
if Alice and Bob use a universal$_2$ hash function family that consists of linear and surjective functions.

Further, if parameters $l$ and $n$ are sufficiently large, we can also obtain a tight bound on the first term of (\ref{eq:straightforward_division_two_regions})
{\it without} relying on the normal approximation of $P_{\rm hg}(c|k)$.
\begin{Lmm}
\label{lmm:upperbound_simple_bound}
If $\frac54s(\varepsilon)^2\le l\le n$, $1\le k$, and $c_{\max}\le0.12l$, we have
\begin{equation}
\sum_{c=0}^{\min(\lfloor\bar{c}-s\sigma\rfloor,c_{\max})}P_{\rm hg}(c|k)
\le
\sqrt{\frac{n+l}{n}}\sqrt{\frac{s(\varepsilon)^2+2\pi}{2}}\,e^{\mu}\varepsilon,
\end{equation}
where $\mu:=1/(6n)+1/(12)$.
Note that this bound holds rigorously, without relying on the normal approximation of $P_{\rm hg}(c|k)$.
\end{Lmm}
This lemma will be proved in \ref{Proof_lemma_simple_bound}.

Now recall that the upper bound $2^{-D+1}$, obtained above for the second term of (\ref{eq:straightforward_division_two_regions}), does not rely on any approximation either.
Hence, besides Proposition 1, we can obtain another bound on $P_{\rm ph}$ that is similarly tight, and is valid rigorously {\it without} relying on any approximation:
\begin{Prop}\label{thm:exact-st_upperbound}
Suppose that $\frac54s(\varepsilon)^2\le l\le n$, and $c_{\max}\le0.12l$
are satisfied for a given $\varepsilon$ (i.e., with $\Phi(s)=\varepsilon$).
Also assume that Alice and Bob perform the QKD protocol specified in Section \ref{sec:protocol_description}.
%
%
Then without using the normal approximation of $P_{\rm hg}(c|k)$, we have
\begin{eqnarray}
P_{\rm ph}
\le \max_k S_{\rm av}(k)
\le 
\sqrt{\frac{s(\varepsilon)^2+2\pi}{2}}\sqrt{\frac{n+l}{n}}\,
e^{\mu}\varepsilon
+2^{-D+1}.
\label{eq:Theorem_3_exact-st_upperbound}
\end{eqnarray}
\end{Prop}

\subsection{The Upper Bounds by The Gaussian Integration}
\label{sec:Upper_Bounds_by_Gaussian_Integration}
In the above analysis of the straightforward bounds, if one wishes to bound $P_{\rm ph}$ by a certain value, say $P_{\max}$, it is necessary to let $D\ge 1-\log_2P_{\max}$.
Hence, if one choose a very small $P_{\max}$ in order to achieve a high level security,
this $D$ can decrease the final key length severely through the sacrificed bit length (\ref{eq:def_alpha_1}).

In this subsection,
we derive improved bounds that holds with $D=1$.
We call them here the {\it Gaussian bounds} for the following reason.
The first step of the analysis is similar to that of the previous section; i.e., we divide the summation of $S_{\rm av}(k)$ as in (\ref{eq:straightforward_division_two_regions2}) and obtain upper bounds for each term.
For the first term of (\ref{eq:straightforward_division_two_regions2}), we use the normal approximation (\ref{eq:def_normal_approx_hypergeo}) again and bound it by $\varepsilon$.
However, for the second term of (\ref{eq:straightforward_division_two_regions2}), we employ a quite different strategy:
We approximate $P_{\rm hg}(k|c)$ by using (\ref{eq:def_normal_approx_hypergeo}), and also upper bound $S_{\rm pa}(k,c)$ by an exponential function of a simple linear function of $c$ (specified below in (\ref{eq:inequality_upper_bound_S_pa_by_beta})).
By using this simple form, we evaluate the summation over $c$ as a Gaussian integral.
As a result of this integration, instead of $2^{-D+1}$ appearing in the previous subsection,
we obtain an upper bound $\delta\varepsilon$ on the second term, with $\delta$ being small for large $l,n$.

In order for this strategy using the Gaussian integration to work properly, parameter $k$ must be confined to a specific region.
Thus as a preparation, we consider the following three cases depending on the value of $k$:
\begin{enumerate}
\item If $k$ is too small (i.e., $0\le k\le nc_{\min}/l$), it can be shown that $S_{\rm pa}(k,c)$ is always bounded by $\varepsilon$, by using the properties of $g(k,c)$.
Thus $S_{\rm av}(k)\le\varepsilon$.
\item For the intermediate domain where $nc_{\min}/l\le k\le (n+l)\hat{p}_{{\rm sft},\varepsilon}(c_{\max})$, the function $g(k,c)$ (used for $S_{\rm pa}(k,c)=2^{[g(k,c)]^-+1}$) can be bounded from above by a simple function, i.e., a constant or a linear function of $c$.
\item If $k$ is too large (i.e., $(n+l)\hat{p}_{{\rm sft},\varepsilon}(c_{\max})\le k$), we can also show that $S_{\rm av}(k)$ is less than $\sum_{c=0}^{\bar{c}-s\sigma}P_{\rm hg}(c|k)$.
\end{enumerate}
The more precise argument will be given in \ref{sec:proof_theorem_1}, and we have the following theorem.
\begin{Thm}
\label{thm:the_first_theorem}
Let $D=1$.
If $c_{\min}\le c_{\max}$ and $2\le s(\varepsilon)$,
then $S_{\rm av}(k)$ is bounded from above as follows 
\begin{itemize}
\item (Case 1) If $0\le k\le nc_{\min}/l$,
\begin{equation}
S_{\rm av}(k)
\le \varepsilon.
\label{eq:case1_upperbound}
\end{equation}
\item (Case 2) If $nc_{\min}/l< k\le (n+l)\hat{p}_{{\rm sft},\varepsilon}(c_{\max})$,
for an arbitrary possible outcome $c$, we have
\begin{equation}
S_{\rm pa}(k,c)\le 
\min \left(2^{-\beta\left(c-(\bar{c}-s\sigma+1)\right)},\ 1\right),
\label{eq:inequality_upper_bound_S_pa_by_beta}
\end{equation}
where
\begin{equation}
\beta:= \frac1{1+4\gamma}\frac{n+l}{l}h'(\hat{p}_{{\rm sft},\varepsilon}(c_{\max})).
\label{eq:def_beta}
\end{equation}
Thus
\begin{eqnarray}
S_{\rm av}(k)
&\le&
\sum_{c=0}^{\min(\lfloor\bar{c}-s\sigma\rfloor,c_{\max})} P_{\rm hg}(c|k)\nonumber\\
&&\ \ \ \ +\sum_{c=\lfloor\bar{c}-s\sigma\rfloor+1}^{c_{\rm max}}
P_{\rm hg}(c|k)
2^{-\beta\left(c-(\bar{c}-s\sigma)+1\right)}.
\label{eq:case2_upperbound}
\end{eqnarray}

\item (Case 3) If $(n+l)\hat{p}_{{\rm sft},\varepsilon}(c_{\max})\le k$,
then $c_{\max}\le \bar{c}-s\sigma$ holds by the definition of $\hat{p}_{{\rm sft},\varepsilon}(c)$.
Hence
\begin{equation}
S_{\rm av}(k)
\le
\sum_{c=0}^{c_{\max}} P_{\rm hg}(c|k)\le\sum_{c=0}
^{\lfloor\bar{c}-s\sigma\rfloor}
P_{\rm hg}(c|k).
\end{equation}
\end{itemize}
\end{Thm}
(For the proof of this theorem, see \ref{sec:proof_theorem_1}.)
We stress that the normal approximation to $P_{\rm hg}(c|k)$ is not yet applied,
and thus all inequalities are rigorous at this stage%
\footnote{It is true that we used the normal approximation in deriving $\hat{p}_{{\rm sft},\varepsilon}(c)$ in (\ref{eq:def_hat_p_sft}) and (\ref{eq:def_hat_p}), and that $\hat{p}_{{\rm sft},\varepsilon}(c)$ is used in the statement of Theorem \ref{thm:the_first_theorem}.
However, in the proof of Theorem 1 we use no approximation; thus the theorem holds rigorously, without any approximation.}

Then in the rest of this subsection, we will show that the right hand side of each inequality of Theorem \ref{thm:the_first_theorem} can be bounded from above by $(1+\delta)\varepsilon$, with $\delta$ being smaller than one for sufficiently large $l,n$.
In other words, we obtain an upper bound on $S_{\rm av}(k)$ that is valid for all $k$;
and thus an upper bound on $P_{\rm ph}$ (recall the argument of Section \ref{sec:upper_bound_Pph_each_k}).
can be bounded from above by (and thus $P_{\rm ph}$) from above by $\varepsilon$.
Let us first discuss the easier cases, namely, Cases 1 and 3.
As mentioned above, for these two cases $S_{\rm av}(k)$ can be easily shown to be less than $\varepsilon$:
For Case 1, it is already proved in Theorem \ref{thm:the_first_theorem}.
For Case 3, if one applies the normal approximation to $P_{\rm hg}(c|k)$,
$S_{\rm av}(k)$ is bounded by $\varepsilon$, as can be seen by the same argument as in the previous section (see the paragraph of (\ref{eq:def_normal_approx_hypergeo})).

Hence it remains to evaluate Case 2, where parameter $k$ is restricted as $nc_{\min}/l< k\le (n+l)\hat{p}_{{\rm sft},\varepsilon}(c_{\max})$.
As mentioned above, we here show that $S_{\rm av}(k)$ can be rewritten as the Gaussian integration in this case.
In Inequality (\ref{eq:case2_upperbound}), the first term on the right hand side can be bounded by $\varepsilon$, with the approximation applied to $P_{\rm hg}(c|k)$.
For the second term, which is a summation over $c$,
we replace $P_{\rm hg}(c|k)$ with the the normal approximation.
In addition to that, we replace $S_{\rm pa}(k,c)$ appearing in the same summation by the right hand side of (\ref{eq:inequality_upper_bound_S_pa_by_beta}).
Then the summation can be rewritten a Gaussian integral:
\begin{eqnarray}
\lefteqn{\sum_{c=\lfloor\bar{c}-s\sigma\rfloor}^{c_{\max}}
P_{\rm hg}(c|k)
2^{-\beta\left(c-(\bar{c}-s\sigma)+1\right)}}\\
&\simeq&
\frac1{\sqrt{2\pi}}\int_{-s}^{(c_{\max}-\bar{c})/\sigma}
\exp\left[-\frac{x^2}2-s\left(x+s\right)\xi_{\varepsilon}(k)\right]dx.\nonumber\\
&\le& \frac1{\sqrt{2\pi}}\int_{-s}^{\infty}
\exp\left[-\frac{x^2}2-s\left(x+s\right)\xi_{\varepsilon}(k)\right]dx.\\
&=&
e^{\frac12{\xi_{\varepsilon}}(\xi_{\varepsilon}-2)s^2}
\frac1{\sqrt{2\pi}}\int_{(\xi_{\varepsilon}-1)s}^\infty
e^{-x^2/2}dx\\
&=:&I_2\left(\xi_{\varepsilon}(k)\right),\nonumber
\end{eqnarray}
where 
\[
\xi_{\varepsilon}(k):=(\ln2)\beta\sigma(k)/s(\varepsilon).
\]

Further, in order to bound $I_2\left(\xi_{\varepsilon}(k)\right)$ using $\varepsilon$, we introduce the inequalities
\begin{equation}
\frac{\sqrt2}{\sqrt{x^2+2\pi}}e^{-x^2/2}\le \Phi(x)\le \frac{\sqrt2}{x}e^{-x^2/2},
\label{eq:inequality_Phi_both_sides}
\end{equation}
where $\Phi(x)$ is the normal distribution function given in (\ref{eq:def_normal_distribution_function}).
(Inequalities (\ref{eq:inequality_Phi_both_sides}) will also be proved in \ref{sec:proof_theorem_1}.)
By using (\ref{eq:inequality_Phi_both_sides}), the integral $I_2\left(\xi_{\varepsilon}(k)\right)$ can be evaluated further as
\begin{equation}
I_2\left(\xi_{\varepsilon}(k)\right) 
\le\frac{\sqrt{1+2\pi s^{-2}}}{\xi_{\varepsilon}(k)-1}\Phi(s(\varepsilon))
=\frac{\sqrt{1+2\pi s^{-2}}}{\xi_{\varepsilon}(k)-1}\,\varepsilon.
\label{eq:I_2_bound_by_xi}
\end{equation}
Note here that $\sigma(k)$ is an increasing function of $k$, because $\xi_{\varepsilon}(k)$ is.
Thus the final term of (\ref{eq:I_2_bound_by_xi}) is maximized at the lower boundary $k=nc_{\min}/l$,
and we obtain finally
\begin{equation}
I_2\left(\xi_{\varepsilon}(k)\right)\le\frac{\sqrt{1+2\pi s^{-2}}}{\xi_{\min,\varepsilon}-1}\,\varepsilon
\label{eq:upper_bound_I_2}
\end{equation}
with $\xi_{\min,\varepsilon}:=\xi_{\varepsilon}(nc_{\min}/l)$.
We now have the following theorem:

\begin{Thm}\label{thm:gauss_upperbound}
For a given $\varepsilon$, 
suppose that $c_{\min}\le c_{\max}$, $2\le s(\varepsilon)$ and $1<\xi_{\min,\varepsilon}$ with
\begin{eqnarray}
\xi_{\min,\varepsilon}&:=&\xi_{\varepsilon}(nc_{\min}/l)\label{eq:def_alpha}\\
&=&\frac{(n+l)\ln2}{s(\varepsilon)l(1+4\gamma)}h'\left(\hat{p}_{{\rm sft},\varepsilon}(c_{\max})\right)\sigma(nc_{\min}/l).\nonumber
\end{eqnarray}
Here $\hat{p}_{{\rm sft},\varepsilon}(c)$ is defined in Eq. (\ref{eq:def_hat_p_sft}), $\sigma$ in Eq. (\ref{eq:def_barc_sigma}), and $h'(x)=\log_2\left(\frac{1-x}{x}\right)$.
Also assume that Alice and Bob perform the QKD protocol specified in Section \ref{sec:protocol_description}.
%
%
Then with the normal approximation applied to $P_{\rm hg}(c|k)$, 
$ P_{\rm ph}$ can be bounded as
\begin{equation}
P_{\rm ph}
\le \max_k S_{\rm av}(k)
\le
(1+\delta)\varepsilon,
\label{eq:P_ph_upper_bound_by_delta}
\end{equation}
where
\begin{equation}
\delta := \frac{\sqrt{1+2\pi s(\varepsilon)^{-2}}}{\xi_{\min,\varepsilon}-1}.
\end{equation}
\end{Thm}

Note here that none of $c_{\min}$, $\hat{p}_{{\rm sft},\varepsilon}(c_{\max})$ or $\gamma$ depends on $k$ or $c$, which can vary for each run of the protocol;
thus $\xi_{\min,\varepsilon}$ can be calculated as a fixed value specified by the protocol. (In other words, $\xi_{\min,\varepsilon}$ is the constant and thus calculated at the preparation stage prior to the protocol.)

Further, as we have done in the previous subsection, if parameters $l$ and $n$ are sufficiently large, we can also obtain a similarly good bound {\it without} relying on the normal approximation of $P_{\rm hg}(c|k)$ (in Eq. (\ref{eq:def_normal_approx_hypergeo})).
By using exact upper bounds on $P_{\rm hg}(c|k)$ including Lemma \ref{lmm:upperbound_simple_bound}, we obtain the following theorem:
\begin{Thm}\label{thm:exact_upperbound}
Suppose that $1\le l\le n$, $s^2\le c_{\min}\le c_{\max}\le 0.12l$, and $1<\xi_{\min}$ are satisfied
for a given $\varepsilon$.
Also assume that Alice and Bob perform the QKD protocol specified in Section \ref{sec:protocol_description}.
%
%
Then {\it without} using the normal approximation of $P_{\rm hg}(c|k)$, we have
\begin{eqnarray}
 P_{\rm ph}
\le \max_k S_{\rm av}(k)
\le 
 P_{{\rm ph},\varepsilon}(c_{\min},\xi_{\min,\varepsilon}),
\end{eqnarray}
where
\begin{eqnarray}
 P_{{\rm ph},\varepsilon}(c_{\min},\xi_{\min,\varepsilon})
&:=&
\sqrt{\frac{s(\varepsilon)^2+2\pi}{2}}\sqrt{\frac{n+l}{n}}e^{\mu}\varepsilon\label{eq:Theorem_3_exact_upperbound}\\
&&+\left(\frac{\sqrt{1+2\pi s(\varepsilon)^{-2}}}{\xi_{\min,\varepsilon}-1}\frac{e^{\mu+\nu}}{\sqrt{1-\frac{s(\varepsilon)}{\sqrt{c_{\min}}}}}+\varepsilon\right)\varepsilon\nonumber,
\end{eqnarray}
where $\mu=1/(6n)+1/12$, $\nu=1/(12l)+1/(2(n+l-1))$.
\end{Thm}
The proof of this theorem is given in \ref{sec:proof_theorem_3}.

\subsection{Second Order Asymptotics}
Now, we roughly estimate the relation between
the sacrifice bit length and 
the upper bound $\max_{k}S_{\rm av}(k)$ of the phase error.
For this purpose, we focus on the asymptotic expansion
for the sacrifice bit.
In the protocol discussed in the above,
the sacrifice bit length 
$\alpha(c)$ is
$\lceil nh\left(\hat{p}_{{\rm sft},\varepsilon}(c+1)\right)\rceil+2$
with
$\hat{p}_{{\rm sft},\varepsilon}(c)=
\frac{(n+l)\hat{p}_\varepsilon(c)-lp_{\rm smp}(c)}n$
and
$\hat{p}_{\varepsilon}(c) := \frac1{1+4\gamma}
\left(
p_{\rm smp}
+2\gamma
+2\sqrt{\gamma\left\{p_{\rm smp}\left(1-p_{\rm smp}\right)+\gamma\right\}}
\right)$.
When the ratio $l/n$ is $t$,
we obtain the asymptotic expansion:
\begin{eqnarray}
\lceil nh\left(\hat{p}_{{\rm sft},\varepsilon}(c+1)\right)\rceil+2
=
nh \left(p_{\rm smp}(c_{\min})\right) + \sqrt{n} 
g_t(p_{\rm smp}(c_{\min})) +o(\sqrt{n}),
\end{eqnarray}
where
$g_t(x):=h' \left(x \right) \sqrt{\frac{x(1-x) (1+t)}{4t}} s(\varepsilon) $.
When we use only the first term in the above expansion, 
the upper bound $\max_{k}S_{\rm av}(k)$ for the phase error 
converges to zero or one.
The limit value zero or one cannot be used for the approximation for 
the upper bound $\max_{k}S_{\rm av}(k)$
because the real value of the upper bound $\max_{k}S_{\rm av}(k)$
takes a value between zero and one, which is different from zero or one.

However, when we use up to the second order $\sqrt{n}$
in the asymptotic expansion of $\alpha(c)$,
the upper bound $\max_{k}S_{\rm av}(k)$
converges to a value between zero and one.
In this case,
we can use the limit for the approximation for
the upper bound $\max_{k}S_{\rm av}(k)$.
That is,
by using the above asymptotic expansion,
the virtual phase error can be abounded as the following way.

\begin{Thm}\label{thm:gauss_asymptotic}
For a given $\varepsilon$, $p_{\min}$, and $p_{\max}$,
we choose $c_{\min}$ and $c_{\max}$ as $p_{\min}l$ and $p_{\max}l$,
and assume that $l/n=t$.
Also suppose that 
Alice and Bob perform the QKD protocol
specified in Section 2, except that the sacrifice bit length $\alpha(c)$ is less than
$nh \left(p_{\rm smp}(c_{\min})\right) + \sqrt{n} g_t(p_{\rm smp}(c_{\min})) $
for $c \in [c_{\min},c_{\max}]$.
Then, the maximum $P_{{\rm ph},n,l}$ of 
$S_{\rm av}(k)$ with given $n$ and $t$
can be asymptotically characterized as
\begin{equation}
\lim_{n \to \infty} \max_{l:l \ge tn } P_{{\rm ph},n,l} \le \varepsilon.
\end{equation}
\end{Thm}

The proof will be given in \ref{proof-gauss_asymptotic}.

\section{How to use the above formulas to evaluate the security of one's QKD system}
\label{sec:how_to_use}
In this section we summarize what we have proved so far, and then explain how one can use Proposition 1 or 2, or Theorem 2 or 3 to evaluate the security of one's QKD system.

\subsection{Summary of Our Results}
As discussed in Section \ref{sec:security_criteria}, the standard quantitative measure of the security of QKD is the trace distance $\left\|\rho_{A,E'}-\rho_{\rm Ideal}\right\|_1$
between the actual state $\rho_{A,E'}$ and the ideal state $\rho_{\rm Ideal}$, given in (\ref{eq:def_uniersal_composability}).
Inequalities (\ref{eq:trace_distance_less_than_Pph_and_exp_alpha}) and (\ref{eq:trace_distance_less_than_Pph}) claim that this trace distance can be bounded from above by the averaged decoding error probability $ P_{\rm ph}$ of the virtual phase error correction.
Throughout the paper, we are interested in bounding $P_{\rm ph}$ by using the Shor-Preskill's formalism.
Also in Section \ref{sec:security_criteria},
we have shown that in order to bound $P_{\rm ph}$ under an arbitrary attack by Eve, it suffices to bound the probability 
$\max_kS_{\rm av}(k)$, with $S_{\rm av}(k)$ defined in (\ref{eq:S_av_bound_1}) (or equivalently, for all $k$, one needs to bound $S_{\rm av}(k)$ by a certain value).
Here the function $S_{\rm av}(k)$ gives an upper bound on the failure probability $S_{\rm pa}(k,c)$ of the virtual phase error correction, averaged with respect to the hypergeometric distribution $P_{\rm hg}(c|k)$.
Our analyses of Sections \ref{sec:upper_confidence_limit} and \ref{sec:upper_decoding_error_probability} are devoted for obtaining an upper bound on $\max_k S_{\rm av}(k)$.

In Section \ref{sec:upper_confidence_limit}, we determined the suitable functional form of the upper bound $\hat{p}_{\rm sft}(c)$ on the phase error rate $\hat{p}_{\rm sft}(k,c)$ of the sifted key, such that we can achieve high key generation rates and the high-level security simultaneously.
The function $\hat{p}_{\rm sft}(c)$ is used for calculating the sacrifice bit length $\alpha(c)$ of Eq. (\ref{eq:def_alpha_1}), i.e., the number of bits that needs to be erased in privacy amplification (PA).
This problem can be reduced to determining an upper bound on parameter $k$, or equivalently, that on the phase error rate  $p_{\rm sft}(k,c)$ of a sifted key.
For this purpose, we derived an upper bound $\hat{p}_{{\rm sft},\varepsilon}(c)$ of Eqs. (\ref{eq:def_hat_p}) and (\ref{eq:def_hat_p_sft}) on $p_{\rm sft}(k,c)$, as a function of the measured error rate $p_{\rm smp}(c)=c/l$ of sample bits.
We here used the standard method of interval estimation, and the upper bound $\hat{p}_{{\rm sft},\varepsilon}(c)$ is defined so that, for any value of $k$, the undesired case $p_{\rm sft}(k,c)>\hat{p}_{{\rm sft},\varepsilon}(c)$ occurs with a probability $\le\varepsilon$ (see Eqs. (\ref{eq:estimation_necessary_condition}) and (\ref{eq:condition_Pr_k_approx})).

Then in Section \ref{sec:upper_decoding_error_probability}, by using this $\hat{p}_{{\rm sft},\varepsilon}(c)$ and the corresponding sacrificed bit length $\alpha(c)$ given in (\ref{eq:def_alpha_1}),
we obtained the upper bounds on $S_{\rm av}(k)$ that holds for all $k$.
By the argument of the paragraph of (\ref{eq:S_av_bound_1}), this means that we have given upper bounds on $P_{\rm ph}$.
For the sake of simplicity, we first gave straightforward bounds in Proposition 1 (with the approximated values of the hypergeometric distribution $P_{\rm hg}(c|k)$) and Proposition 2 (without any approximation).
Next we gave the other bounds exploiting the properties of  the Gaussian integration, which yield larger final key length $G$ for sufficiently large $l,n$;
namely, Theorem 2 (with the approximated $P_{\rm hg}(c|k)$) and Theorem 3 (without any approximation).

\subsection{How to Use The Straightforward Upper Bounds}
\subsubsection{The Straightforward Upper Bound With The Normal Approximation (How to Use Proposition 1)}
\label{sec:how_to_use_proposition1}
Here we present how to calculate the secret key length of one's QKD system using the straightforward upper bound on $P_{\rm ph}$ obtained in Propositions 1.
\setcounter{footnote}{0}
\newcounter{itemcounter}
\begin{itemize}
\item Preparation steps:
\begin{enumerate}
\item Determine one's desired upper bound $T_{\max}$ on trace distance.
\item Calculate the corresponding upper bound on the phase error rate by $P_{\max}=\frac18(T_{\max})^2$.
\item \label{item:straightforward_choose_s}
Let the confidence limit be $\varepsilon=\frac12P_{\max}$.
Then calculate parameter $s=\Phi^{-1}(\varepsilon)$,
as the inverse value of  the normal distribution function $\Phi(x)$
(see the definitions of $\Phi(x)$ and $s(\varepsilon)$ given in (\ref{eq:def_normal_distribution_function}), (\ref{eq:def_s_epsilon})).
\item Let $D=\lceil 2-\log_2P_{\max} \rceil$.
\item Determine $c_{\min}$ and $c_{\max}$.
\item \label{item:straightforward_set_epsilon}
(Parameter check:) No parameter check is necessary for Proposition 1.
\setcounter{itemcounter}{\value{enumi}}
\end{enumerate}
\end{itemize}
Under this setting of parameters, one can guarantee that $P_{\rm ph}\le \varepsilon+2^{-D+1}\le P_{\max}$, by applying the normal approximation to $P_{\rm hg}(c|k)$ and by using Proposition 1.
Then Inequality (\ref{eq:trace_distance_less_than_Pph}) guarantees that the trace distance is bounded as $\|\rho_{A,E'}-\rho_{\Ideal} \|_1\le 2\sqrt{2}\sqrt{P_{\max}}=T_{\max}$.
(As specified below, we here assume that Alice and Bob use a universal$_2$ hash function family that consists of linear and surjective functions.)

\begin{itemize}
\item For each run of the protocol:
\begin{enumerate}
\setcounter{enumi}{\value{itemcounter}}
\item
\label{step:each_run_of_the_protocol}
Perform the protocol as specified in Section \ref{sec:protocol_description}.
In particular in the PA step, for the calculation of the length $\alpha(c)$ of (\ref{eq:def_alpha_1}),
use $\hat{p}_{{\rm sft},\varepsilon}(c)$ defined in Eqs. (\ref{eq:def_hat_p}) and (\ref{eq:def_hat_p_sft}), 
as well as parameters $s$ and $D$ obtained in the preparation steps above.%
\footnote{Throughout this section, we neglect the deviation of $l,n$ from their averages when the bases $x,z$ are chosen with a constant probability, and assume that they are constant.}
Then use a universal$_2$ hash function family that consists of linear and surjective functions, to convert the reconciled key to the secret key.
\end{enumerate}
\end{itemize}
As noted in Section \ref{sec:protocol_description},
as a result of this protocol, Alice and Bob obtain the final key of length $G=n_{\rm rec}-\alpha(c)$ with $\alpha(c)$ given in (\ref{eq:def_alpha_1}), and
$n_{\rm rec}$ being the reconciled key length.
If an error correcting code with efficiency $f$ is used, we have $n_{\rm rec}=n(1-fh(p_{\rm bit}))$, with $p_{\rm bit}$ being the bit error rate of the sifted key.
Thus Alice and Bob obtain the final key of length $G$, given in (\ref{eq:def_G_c}).

\subsubsection{The Straightforward Upper Bound Without Any Approximation (How to Use Proposition 2)}
\label{sec:how_to_use_proposition2}
By using Proposition 2, an exact upper bound on $P_{\rm ph}$ can be obtained, without relying on the normal approximation of $P_{\rm hg}(c|k)$.
In this case all the steps are the same as those given in Section \ref{sec:how_to_use_proposition1}, except for Steps (\ref{item:straightforward_choose_s}) and (\ref{item:straightforward_set_epsilon}):
\begin{itemize}
\item[(\ref{item:straightforward_choose_s}')]
Choose parameter $s$ such that
\[
\sqrt{\frac{n+l}{n}}\sqrt{\frac{s^2+2\pi}{2}}\,e^{\mu}\Phi(s)\le \frac12P_{\rm max}
\]
is satified, where  $\mu=1/(6n)+1/12$.
\item[(\ref{item:straightforward_set_epsilon}')]
(Parameter check:)
Check that $\frac54s^2\le l\le n$ and $c_{\max}\le0.12l$ are satisfied. If not, set $T_{\max}$ smaller and restart from Step (i).
\end{itemize}

As a result of Step (\ref{item:straightforward_choose_s}'), we have $\varepsilon =\Phi(s(\varepsilon)) \le s^{-1}\times \frac12P_{\rm max}$.
This means that, for a fixed value of $P_{\rm max}$,
one needs to choose $\varepsilon=\Phi(s(\varepsilon))$ to be smaller than that obtained in Section \ref{sec:how_to_use_proposition1}, by a factor of $s^{-1}$.
As a result, $s$ also turns out to be larger, one ends up with a smaller final key length.
Note, however, that such increment of $s$ is negligible for sufficiently large $s$ (e.g., for $s\ge10$), because $\Phi(s)$ scales as $e^{-\frac12s^2}$ and thus a very small increment of $s$ compensates the factor of $s^{-1}$ in front of $\frac12P_{\rm max}$.
Hence the decrement in the final key length is very small.
We will demonstrate this fact in the next section by a numerical calculation in Section \ref{sec:exact_vs_approximate}.

\subsection{How to Use The Upper Bounds by The Gaussian Integration (How to Use Theorems 2 and 3)}
\label{sec:how_to_use_gaussian_bounds}
As mentioned in Section \ref{sec:Upper_Bounds_by_Gaussian_Integration},
if parameters $l$ and $n$ are sufficiently large,
we can set $D=1$ and still obtain similarly tight bounds on $P_{\rm ph}$ as given in Theorems 2 and 3;
thereby we can improve the final key length $G$.
For these cases too, we summarize how to calculate the secret key length of one's QKD system.

\subsubsection{The Gaussian Bound With The Normal Approximation (How to Use The Bound of Theorem 2)}
\label{sec:gaussian_bound_without_approximation}


For Theorem 2, the preparation steps are modified as follows:
\begin{itemize}
\item Preparation steps:
\begin{enumerate}
\item Determine one's desired upper bound on trace distance $T_{\max}$.
\item Calculate the corresponding upper bound on the phase error rate by $P_{\rm max}=\frac18(T_{\max})^2$.
\item
\label{item:set_epsilon_smaller}
Set the confidence limit $\varepsilon$ to be slightly smaller than $P_{\rm max}$.
(For example, if $l,n$ are sufficiently large, $\varepsilon=0.9P_{\rm ph}$ is usually sufficient.)
Then calculate parameter $s=\Phi^{-1}(\varepsilon)$,
as the inverse value of  the normal distribution function $\Phi(x)$ given in (\ref{eq:def_normal_distribution_function}).
\item Let $D=1$.
\item Determine $c_{\min}$ and $c_{\max}$, such that the conditions in the first sentence of Theorem 2 are all satisfied.
\label{item:step_before_final}
\item (Parameter Check:) Check if $\delta$ is small enough so that Inequality (\ref{eq:P_ph_upper_bound_by_delta}) is satisfied.
If not, go back to Step (\ref{item:set_epsilon_smaller}) and set $\varepsilon$ smaller.
\label{item:final_step}
\setcounter{itemcounter}{\value{enumi}}
\end{enumerate}
\end{itemize}

After these preparation steps, Alice and Bob run the protocol as in previous sections.
That is, they run the protocol as specified in Step (\ref{step:each_run_of_the_protocol}) of Section \ref {sec:how_to_use_proposition1}.


\subsubsection{The Gaussian Bound Without The Normal Approximation (How to Use The Bound of Theorem \ref{thm:exact_upperbound})}
\label{subsec-exact}
As we have done for the case of the straightforward bounds,
we also obtained in Theorem \ref{thm:exact_upperbound} the exact version of the Gaussian bound that does not rely on the normal approximation of $P_{\rm hg}(c|k)$.
This theorem was derived using essentially the same idea as Theorem 2 and achieves a similarly tight bound, but it does not rely on any approximation.


For Theorem \ref{thm:exact_upperbound},
the preparation steps are the same as Theorem 2 (i.e., the same as in Section \ref{sec:gaussian_bound_without_approximation}),
except for Steps (\ref{item:step_before_final}) and (\ref{item:final_step}):
\begin{itemize}
\item[(\ref{item:step_before_final}'')] Determine $c_{\min}$ and $c_{\max}$, such that the conditions in the first sentence of Theorem \ref{thm:exact_upperbound} are all satisfied.
\item[(\ref{item:final_step}'')] (Parameter Check:) Check if $\delta'$ is small enough so that Inequality (\ref{eq:Theorem_3_exact_upperbound}) is satisfied.
If not, go back to Step (\ref{item:set_epsilon_smaller}) and set $\varepsilon$ smaller.
\end{itemize}

After these preparation steps, Alice and Bob run the protocol as in previous sections.
That is, they run the protocol as specified in Step (\ref{step:each_run_of_the_protocol}) of Section \ref {sec:how_to_use_proposition1}.

\subsection{Rough Estimate of The Key Rate and The Security Parameter}
We note here that if $l,n$ are sufficiently large, parameters $\gamma$ and $\delta$ becomes sufficiently small, and the approximate evaluation of the key length $G$ of (\ref{eq:def_G_c}) can be greatly simplified.

As one can see from Steps (i) and (ii) of Section \ref{sec:how_to_use_gaussian_bounds}, bounding $P_{\rm ph}$ is enough for the security.
If $\delta$ is sufficiently small, then according to Theorem 2 (or or Step (iii) of Section \ref{sec:how_to_use_gaussian_bounds}), $P_{\rm ph}$ can be bounded approximately by $\varepsilon$, which determines the value of $\hat{p}_{{\rm sft},\varepsilon}(c)$ via Eqs. (\ref{eq:def_hat_p}) and (\ref{eq:def_hat_p_sft}).
Then as we discussed in the paragraph of Eq. (\ref{eq:approx_hat_p}), if $\gamma$ is sufficiently small, $\hat{p}_{{\rm sft},\varepsilon}(c)=\frac{n+l}{n}\hat{p}_{\varepsilon}(c)-\frac{l}{n}p_{\rm smp}(c)$ can be approximated by using $\hat{p}_{\varepsilon}(c)\simeq p_{\rm smp}(c)+\frac{s}{l}\sigma_{\rm bin}(c)$.

As a result, if the conditions of the first sentence of Theorem 2 are satisfied for a given set of experimental parameters, and if $\gamma$ and $\delta$ are sufficiently small, one has the following rough estimates.
The trace distance is approximately bounded by the square root of $\varepsilon$ as
\begin{eqnarray*}
||\rho_{A,E}-\rho_{\rm Ideal}||&\le& 2\sqrt2\sqrt{ P_{\rm ph}},\\
 P_{\rm ph}\le(1+\delta)\varepsilon&\simeq&\varepsilon.
\end{eqnarray*}
Parameter $s$ is chosen to be the deviation of the standard deviation, i.e., $s=\Phi^{-1}(\varepsilon)$.
Then this $s$ determines the final key length $G$ as
\begin{eqnarray*}
G&\simeq&n\left[1-fh(p_{\rm bit})-h\left(\hat{p}_{{\rm sft},\varepsilon}(c)\right)\right],\\
\hat{p}_{{\rm sft},\varepsilon}(c)&=&\frac{n+l}{n}\hat{p}_{\varepsilon}(c)-\frac{l}{n}p_{\rm smp}(c),\\
p_{\rm smp}(c)&=&c/l,\\
\hat{p}_{\varepsilon}(c)&\simeq&p_{\rm smp}(c)+\frac{s}{l}\sigma_{\rm bin}(c)\\
&=& p_{\rm smp}(c)+\frac{s}{l}\sqrt{lp_{\rm smp}(c)(1-p_{\rm smp}(c))}.
\end{eqnarray*}
We expect that these relation will be useful for experimentalists and theorists who wish to obtain a rough estimate of the key length with the finite size effect taken into account.


\section{Numerical results.}
\label{sec:numerical_results}
We demonstrate the tightness of our bound with numerical results.
We consider a quantum channel in the absence of eavesdropper, and assume that it can be described as a binary symmetric channel with quantum bit error rate (QBER).

\subsection{Case 1: Basis Choice with Probability $q=\frac12$}
First, as a comparison to preceding literature \cite{SR08,TLGR11}, we plot key rates for the case where Alice and Bob choose the $x$ and the $z$ bases with the equal probability.
We present two types of evaluations given in Section \ref{sec:how_to_use};
one is the analysis of Section \ref{sec:how_to_use_proposition2} using the straightforward bound of Proposition 2,
the other is that of Section \ref{subsec-exact} using the Gaussian bound of Theorem 3. 
Note that both these bounds are derived {\it without} using the normal approximation;
thus the all key generation rates obtained in this subsection are rigorous.

We assume that Alice and Bob choose both the phase basis and the bit basis with probability $q=1/2$, and thus $n=l=N/4$.
We also assume that Alice and Bob consume $r=40$ bits of a previously shared secret key for exchanging the hash value, in order to guarantee that $\epsilon_{\rm cor}\le 10^{-12}$ (in the following, these $r=40$ bits will be subtracted from the final key length $G$).
Then we choose $P_{\max}$ to be $P_{\max}= 0.98\times\frac18\times10^{-20}$,
so that the trace distance $\|\rho_{A,E'}-\rho_{\Ideal} \|_1$ is guaranteed to be less than $T_{\max}=2\sqrt{2P_{\rm \max}}= 0.99\times 10^{-10}$.
By these choices of parameters, we can guarantee $T_{\max}+\epsilon_{\rm cor}\le 10^{-10}$, which is the same condition as used in Ref. \cite{TLGR11}.

Because $r=40$ bits are consumed for guaranteeing that Alice's and Bob's final keys are equal, the effective final key length is $G(c)-r$, with $G(c)$ defined in (\ref{eq:def_G_c}).
Hence in this section, we define the final key rate to be 
\begin{eqnarray}
R(c)&:=&\frac{G(c)-r}{n}\\
&=&\frac1{n}\left[n\left(1-fh(c/l)\right)-\left\lceil nh\left(\hat{p}_{{\rm sft},\varepsilon}(\max\{c,c_{\min}\}+2)\right)\right\rceil-(D+r)\right].\nonumber
\end{eqnarray}
The efficiency of bit error correction is chosen to be $f=1.1$.

\subsubsection{The Straightforward Bound}
\label{sec:numerical_half_straight}
With the above choices of parameters, we perform the analysis of Section \ref{sec:how_to_use_proposition2}, and obtain the corresponding final key rate $R$.
Here we restrict ourselves to the case where parameters $l,n$ satisfy $125\le l=n$.
Parameters $P_{\max}$ and $T_{\max}$ are already specified above.
As to parameter $s$, we follow Step (iii') and let $s=9.9$, so that
\[
\sqrt{\frac{n+l}{n}}\sqrt{\frac{s^2+2\pi}{2}}\,e^{\mu}\Phi(s)\le \sqrt{s^2+2\pi}\,e^{1/4}\Phi(s)
\le1.1\times10^{-22}\le\frac12P_{\rm max}.
\]
According to Step (iv), we choose $D=\lceil 2-\log_2P_{\max}\rceil=79$; next according to Step (v), $c_{\rm min}=0.01l$ and $c_{\max}=0.12l$.
It is easy to verify that all these parameters are compatible with the parameter checks of Step (vi').

Then we assume that Alice and Bob perform the BB84 protocol (i.e., Step (vii)), in the quantum channels with ${\rm QBER}=1\%, 2.5\%$, and $5\%$.
The corresponding key rates $R(c)$ (with $c=l\times {\rm QBER}$) are shown in bold curves in Fig. \ref{key_rate.fig}, versus $n+l$.

\subsubsection{The Gaussian Bound}
\label{sec:numerical_half_gaussian}
\begin{figure}[h]
\centering
\includegraphics[width=15.00cm, clip]{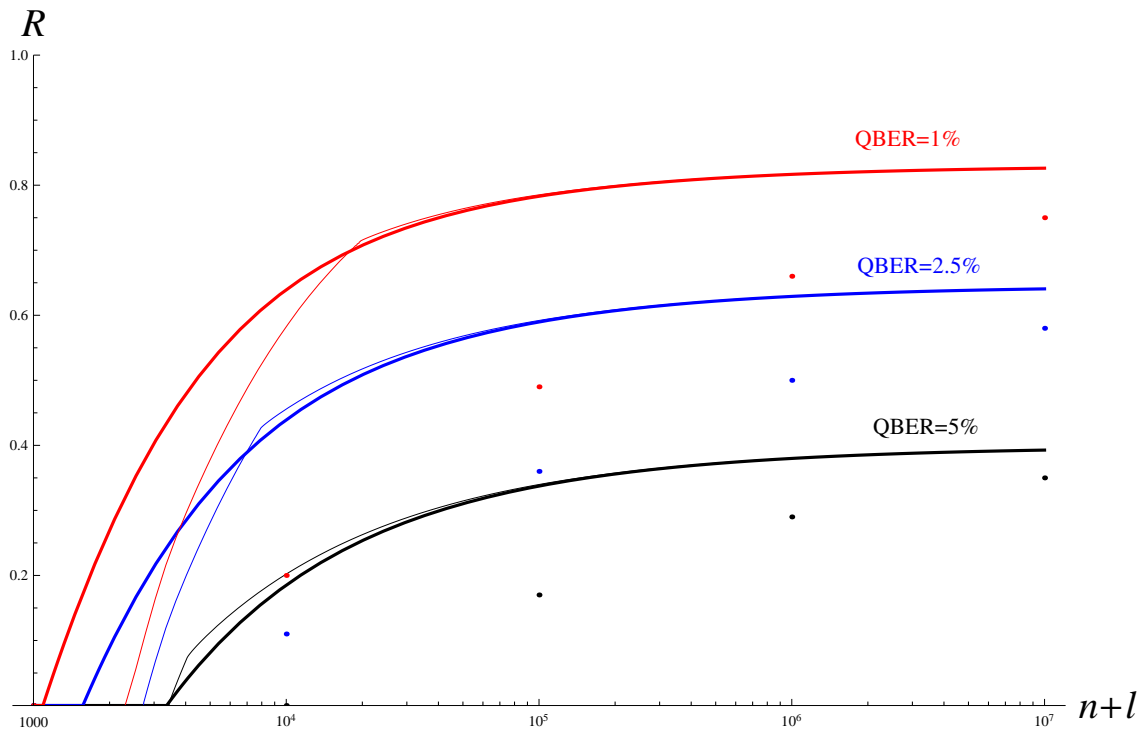}
\caption{(Color online) Key generation rate $R=(G-r)/n$ versus $n+l$, which is the sum of lengths of a sifted key and sample bits.
Here we assume that $x$ and the $z$ bases are chosen with the equal probability, i.e., $q=\frac12$.
The typical QBER are chosen to be 1\% (red), 2.5\% (blue), and 5\% (black).
As to the security, we set $r=40$ and $P_{\rm max}< 0.98\times\frac18\times10^{-20}$, so that $T_{\max}+\epsilon_{\rm corr}\le 10^{-10}$.
That is, the sum of the trace distance and $\epsilon_{\rm cor}$ is less than $10^{-10}$.
We have used two types of analysis to achieve this value of $P_{\rm max}$:
The bold curves represent the key rates based on the straightforward bound given in Proposition 2 and in Section \ref{sec:how_to_use_proposition2}.
The thin curves are based on the Gaussian bound given in Theorem \ref{thm:exact_upperbound} and in Section \ref{subsec-exact}.
We stress that these curves are obtained {\it without} using the normal approximation.
Dots of the same color are the rates obtained in Figure 2 of Ref. \cite{TLGR11}.
} \label{key_rate.fig}
\end{figure}

For the same choice of parameters $q, r, P_{\max},D$,
and for the same ratio of $c_{\max}=0.12l$ with respect to $l$,
we perform the analysis of Section \ref{subsec-exact}.
The remaining parameters to be fixed are $s$ and $c_{\min}$;
hence we here numerically calculate the pairs of $s$ and $c_{\min}$ that gives the best key rate $R(c)$.
That is, we first fix $l$ and $n$, and then search for the pair of $s$ and $c_{\min}$ that is compatible with the parameter check and gives the largest $R(c)$.
(This corresponds to repeating Steps (iii) through (vi') of Section \ref{subsec-exact},
by letting $\varepsilon$ smaller each time, until the largest key length $G(c)$ is obtained.) 
The results are shown in thin curves in Fig. \ref{key_rate.fig}.





As one can see from Fig. \ref{key_rate.fig}, if QBER=5\%,
the Gaussian bound gives better key rate than the straightforward bound for all $l,n$.
On the contrary, for smaller QBER (1\% and 2.5\%), the straightforward bound becomes better for $l,n\simeq 5000$.

The dots in Fig. \ref{key_rate.fig} represents the key rates obtained by Tomamichel et al. \cite{TLGR11} under the same condition.
It can be clearly seen that our key rates $R$ are better in all parameter regions.
For example, Fig. \ref{key_rate.fig} gives $R=0.19$ for ${\rm QBER}=5\%$ and $n+l=10^4$, while Tomamichel et al. gave $R=0$ in this region \cite{TLGR11}.
As $n+l$ becomes larger, $R$ converge very fast to the asymptotic values; all three curves reach more than 80\% of the asymptotic values at $n+l=2\times10^5$.

In particular, as the key size becomes larger, $R$ converge very fast to the asymptotic values, more than 80\% of the asymptotic values at $n+l=2\times10^5$.
As we have noted in Section \ref{sec:protocol_description}, key distillation is quite practical even in this region.
That is, the sizes of bit error correcting codes are independent of security, and thus Alice and Bob may perform bit error correction by dividing a sifted key of $n$ bits to arbitrarily smaller blocks.
As to privacy amplification, one can use the efficient algorithm for the multiplication of the (modified) Toeplitz matrix and a vector.

\subsection{Case 2: Optimized Basis Choice with Variable Probability $q$}
Next, as a more practical setting, we consider the case where Alice and Bob choose the $x$ and the $z$ bases with varying probabilities $q$, $1-q$ (thus, $l=q^2N$, $n=(1-q)^2N$).
Then we maximize the secret fraction $F$, defined by 
\begin{eqnarray}
F(c)&=&\frac{G(c)-r}{N}\\
&=&\frac{1}{N}\left[n\left(1-fh(c/l)\right)-\left\lceil nh\left(\hat{p}_{{\rm sft},\varepsilon}(\max\{c,c_{\min}\}+2)\right)\right\rceil-(D+r)\right]\nonumber
\end{eqnarray}
with respect a fixed raw key length $N$, where $G$ denotes the final key length.
We use the analysis of Section \ref{subsec-exact} based on the Gaussian bound of Theorem 3 (without any approximation);
hence again, all the final key rates obtained in this subsection are rigorous.
We choose parameters $ P_{\rm max}$, $\epsilon_{\rm cor}$ are chosen to be the same as in the previous subsection.
According to Step (iii), we let $s(\varepsilon)=10.5$ so that $\varepsilon= 4.32\times 10^{-26}<\!\!< P_{\rm max}$.
The channel error rates are chosen to be ${\rm QBER}=1\%$, $2.5\%$, and $5\%$, respectively.

\begin{figure}[h]
\centering
\includegraphics[width=15.00cm, clip]{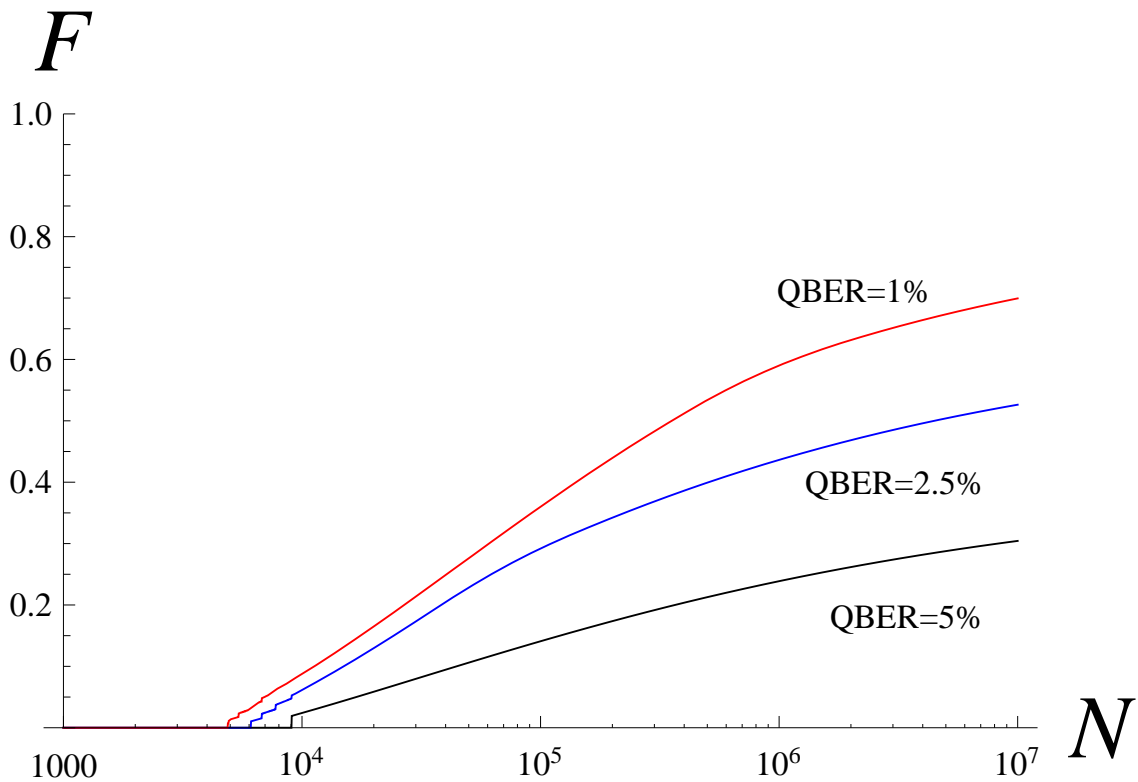}
\caption{(Color online) Secret fraction $F=(G-r)/N$ versus raw key length $N$.
Here we assume that Alice and Bob choose the $x$ and the $z$ bases with varying probabilities $q$, $1-q$.
The probability $q$ and the minimum errors $c_{\rm min}$ are also optimized to give maximum $F$.
The typical QBER are chosen to be 1\% (red), 2.5\% (blue), and 5\% (black).
Parameters $ P_{\rm ph}$, $\epsilon_{\rm cor}$ are chosen to be the same as in Figure \ref{key_rate.fig},
so that  $T_{\max}+\epsilon_{\rm corr}\le 10^{-10}$ is satisfied.
} \label{asymmetric_key_rate.fig}
\end{figure}

Under these settings,  for each fixed value of $N$, we performed numerical simulations to select the optimal values of $q$ and $c_{\min}$ that give the maximum value of $F(c)$.
That is, we first fix $N$, and then search for the pair of $q$ and $c_{\min}$ that is compatible with the parameter check of Step (vi'') and gives the largest $F(c)$.
The results are shown in Figure \ref{asymmetric_key_rate.fig}.

\subsection{Exact Bounds Verses Approximate Bounds}
\label{sec:exact_vs_approximate}
All the key rates of the previous two subsections are rigorous, in the sense that they are obtained without using any approximation.
In this final subsection, we demonstrate that, for practical parameter regions, the key rates are almost the same, whether one uses the analysis based on the normal approximation (i.e., Proposition 1 and Theorem 2), or those without any approximation (i.e., Proposition 2 and Theorem 3).

\begin{figure}[h]
\centering
\includegraphics[width=15.00cm, clip]{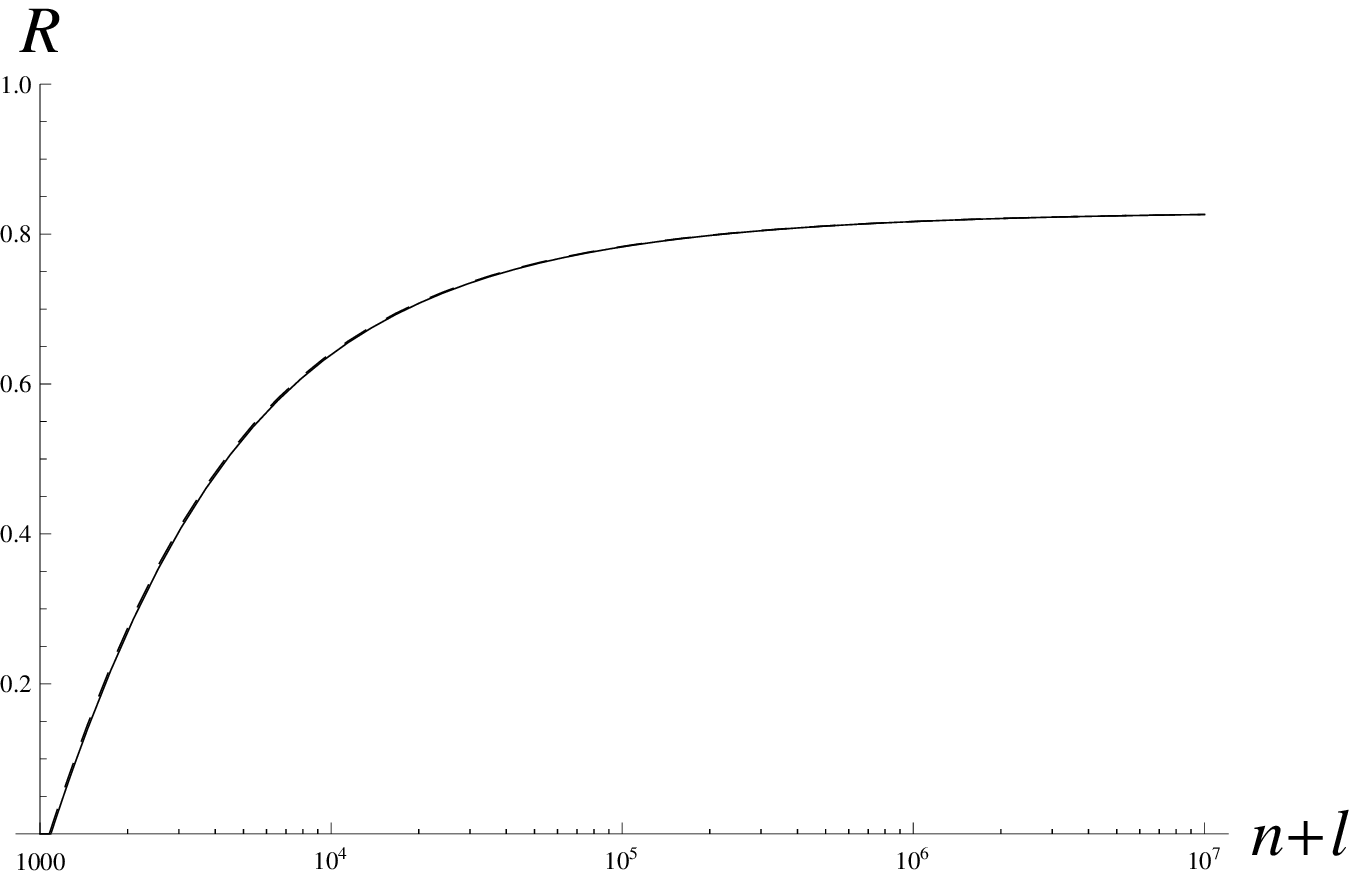}
\caption{
Solid Curve:  the same curve as the solid curve in Figure 1 with QBER=1\%.
This curve is obtained by using Proposition 2, without using any approximation.
Dashed Curve: The final key rate $R(c)$ obtained for the same values of QBER, $P_{\max},r,l,n$, using the straightforward bounds of Proposition 1;
hence this curve is obtained using the normal approximation.
Note that the two curves are almost identical.
}
\label{Exact_vs_Approximate1.fig}
\end{figure}

\begin{figure}[h]
\centering
\includegraphics[width=15.00cm, clip]{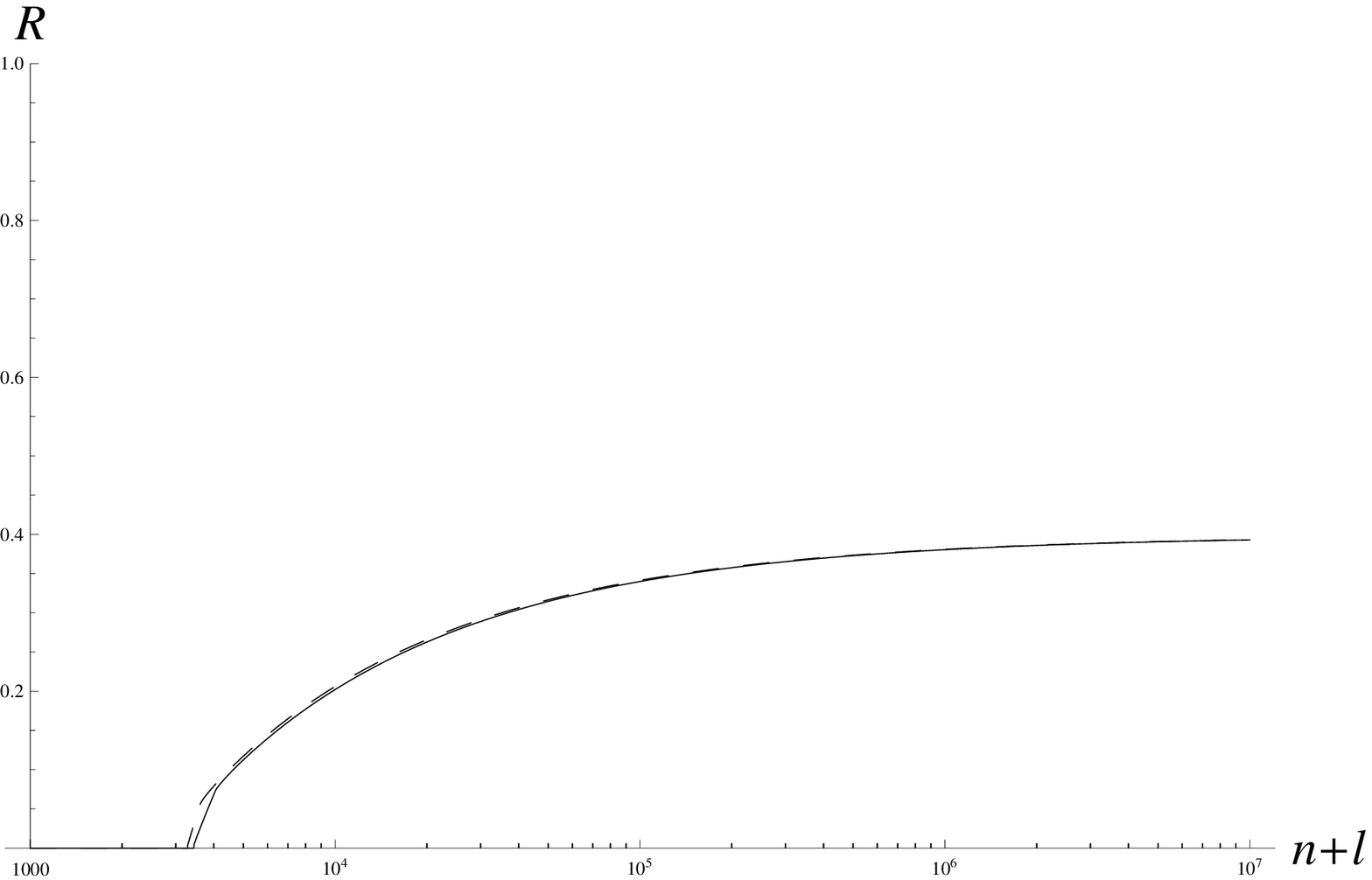}
\caption{
Solid Curve:  the same curve as the thin curve in Figure 1 with QBER=5\%.
This curve is obtained by using Theorem 3 without using any approximation.
Dashed Curve: The final key rate $R(c)$ obtained for the same values of QBER, $P_{\max},r,l,n$, using the straightforward bounds of Theorem 2;
hence this curve is obtained using the normal approximation.
Note again that the two curves are almost identical.
}
\label{Exact_vs_Approximate2.fig}
\end{figure}

In Fig. \ref{Exact_vs_Approximate1.fig},
the solid curve shows $R(c)$ obtained in Section \ref{sec:numerical_half_straight} with QBER=1\%.
On the other hand, the dashed curve in the same figure is the key rate $R(c)$ obtained for the same values of QBER and $P_{\max},r,l,n$ by the procedure of Section \ref{sec:how_to_use_proposition1};
hence this curve is obtained by using Proposition 1, and thus relies on the normal approximation of $P_{\rm hg}$.
Similarly in Fig. \ref{Exact_vs_Approximate2.fig}, 
the solid curve shows $F(c)$ obtained in Section \ref{sec:numerical_half_gaussian} with QBER=5\%,
whereas the dashed curve is obtained by using Theorem 2, which relies on the normal approximation
(Here we performed the optimization of $s$ and $c_{\min}$).

Note that for both of these cases, the exact key rate and approximate key rate are almost identical.
These results suggest that the simple analysis using the normal approximation (i.e., Proposition 1 or Theorem 2) can be justified for the security evaluations of practical QKD systems.

\section{Summary}
In this paper, we presented a concise analysis for the BB84 protocol that takes the finite key effect into account and yields better key generation rates, with and without relying on the normal approximation.
Our results are indeed an improvement of preceding literature;
as we have shown in Figure \ref{key_rate.fig}, our analysis give better key generation rates $R$ in practical settings than in Refs. \cite{SR08,TLGR11}.

In order to serve the convenience of experimentalists who wish to evaluate the security of their QKD systems, we included explicit procedures of security evaluation in Sections \ref{sec:security_criteria} and \ref{sec:how_to_use}.
In particular, in addition to presenting the exact values of key rates and security parameters, we also presented how to obtain their rough estimates using the normal approximation.

For the sake of simplicity, we restricted ourselves to the simple case where Alice has a perfect single photon source.
On the other hand, in order to achieve a long communication distance by a practical QKD system using a weak coherent light source, decoy pulses are necessary \cite{decoy}.
This situation was analyzed by one of the authors \cite{H07}, relying on the normal approximation.
A thorough and exact analysis in this direction without any approximation remains as future work.

\ 

{\bf Acknowledgments}
The authors thank Ryutaroh Matsumoto for valuable comments.
MH is partially supported by a MEXT Grant-in-Aid for Young Scientists (A) No. 20686026 and Grant-in-Aid for Scientific Research (A) No. 23246071.
The Center for Quantum Technologies is funded by the Singapore
Ministry of Education and the National Research Foundation
as part of the Research Centres of Excellence programme.
MH and TT are partially supported by the National Institute of Information and Communication Technolgy (NICT), Japan.

\

\appendix

\section{Justification for the restricting the argument to the generalized Pauli channel}
The generalized Pauli channel is defined to be a channel where the phase error and the bit errors occur stochastically (i.e., with a classical probability).
It is easy to see that, in this setting, the virtual phase error probability $P_{\rm ph}$ after the privacy amplification, mentioned in the main text, can clearly be defined.
In Ref. \cite{H07}, it is shown that the trace distance can be bounded from above by using $P_{\rm ph}$.

Here we demonstrate that, without loss of generality, this argument can be extended to the case where the quantum channel $\Lambda$ between Alice and Bob is arbitrary and general.
First, we consider the discrete twirling.
For $n$-bits sequence $x=(x_1, \ldots, x_n)$ and $z=(x_1, \ldots, z_n)$, define the unitary matrix 
$U(x,z):=(X^{x_1}\otimes X^{x_2}\otimes \cdots\otimes X^{x_n}) (Z^{z_1}\otimes Z^{z_2}\otimes \cdots\otimes Z^{z_n})$,
where $X$ is the bit flip operator and $Z$ the phase flip operator.
Then, the discrete twirling of $\Lambda$
is defined as $\overline{\Lambda}:= \sum_{{\bf z}} 2^{-2n}\Lambda_{{\bf z}}$,
where 
${\bf z}=(x,z)$ and
$\Lambda_{x,z} (\rho):=U(x,z)\Lambda (U(x,z)\rho U(x,z)^\dagger)U(x,z)^\dagger$.
In this paper, 
we treat the phase error and the bit error of the channel $\overline{\Lambda}$ due to the following reason.

Now, we denote the final state 
and the ideal state with the public information $x$
by $\rho_{A,E'|x}(\Lambda)$ and $\rho_{{\rm Ideal}|x}(\Lambda)$
when the channel between Alice and Bob is $\Lambda$.
Hence, our security criterion is 
$\sum_{x} P_{\rm pub}(x)\|\rho_{A,E'|x}(\Lambda)- \rho_{{\rm Ideal}|x}(\Lambda) \|_1$.
Indeed, the distribution $P_{\rm pub}(x)$ depends on the channel $\Lambda$ in general,
however,
it does not change even if the channel is replaced by $\Lambda_{{\bf z}}$
because the initial random variable is uniform and the hash function and error correction are linear.
Also for the same reason, we have
$\|\rho_{A,E'|x}(\Lambda)- \rho_{{\rm Ideal}|x}(\Lambda) \|_1=
\|\rho_{A,E'|x}(\Lambda_{{\bf z}})- \rho_{{\rm Ideal}|x}(\Lambda_{{\bf z}}) \|_1$.
The state 
$\sum_{{\bf z}}2^{-2n}
\rho_{A,E'|x}(\Lambda_{{\bf z}})\otimes |{\bf z}\rangle \langle {\bf z}|$
and
$\sum_{{\bf z}}2^{-2n}
\rho_{{\rm Ideal}|x}(\Lambda_{{\bf z}})\otimes |{\bf z}\rangle \langle {\bf z}|$
can be regarded as the state 
$\rho_{A,E'|x}(\overline{\Lambda})$
and $\rho_{{\rm Ideal}|x}(\overline{\Lambda})$
because the classical information ${\bf z}$ can be treated as a part of Eve's system
with the channel $\overline{\Lambda}$.
Hence, 
\begin{eqnarray*}
&\sum_{x} P_{\rm pub}(x)\|\rho_{A,E'|x}(\Lambda)- \rho_{{\rm Ideal}|x}(\Lambda) \|_1 \\
=&
\|\sum_{{\bf z}}2^{-2n}
\sum_{x} P_{\rm pub}(x)\|\rho_{A,E'|x}(\Lambda_{{\bf z}})\otimes |{\bf z}\rangle \langle {\bf z}|
- \rho_{{\rm Ideal}|x}(\Lambda_{{\bf z}})\otimes |{\bf z}\rangle \langle {\bf z}| \|_1 \\
=&
\|
\sum_{x} P_{\rm pub}(x)\|
\sum_{{\bf z}}2^{-2n}
\rho_{A,E'|x}(\Lambda_{{\bf z}})\otimes |{\bf z}\rangle \langle {\bf z}|
- 
\sum_{{\bf z}}2^{-2n}
\rho_{{\rm Ideal}|x}(\Lambda_{{\bf z}})\otimes |{\bf z}\rangle \langle {\bf z}| \|_1 \\
=&
\sum_{x} P_{\rm pub}(x)\|\rho_{A,E'|x}(\overline{\Lambda})
- \rho_{{\rm Ideal}|x}(\overline{\Lambda}) \|_1.
\end{eqnarray*}
Therefore,
it is enough to consider the case when the channel is $\overline{\Lambda}$
even if the used channel $\Lambda$ is not a Pauli channel.

\section{Proof of Lemma \ref{lmm:upperbound_simple_bound}}
In order to prove this lemma, we introduce several new lemmas.
In the first part, i.e, \ref{sec:uppr_bounds_P_hg}, we derive exact upper bounds on $P_{\rm hg}(c|k)$ given in terms of $l$ or $s(\varepsilon)$.
Then in \ref{sec:upper_and_lowe_bounds_on_Phi}, we show that those upper bounds can also be bounded by $\varepsilon=\Phi^{-1}(s(\varepsilon))$.
Finally in \ref{Proof_lemma_simple_bound}, using the obtained results, we prove Lemma \ref{lmm:upperbound_simple_bound}.

\subsection{Upper Bounds on sums of $P_{\rm hg}(c|k)$}
\label{sec:uppr_bounds_P_hg}
\begin{Lmm}\label{lmm:upperbound_Dlnk}
If $l\le n$ and $\frac{1}{n+l}\le\frac{k}{n+l}\le\frac12$,
\begin{equation}
\sum_{i=0}^{c}P_{\rm hg}(i|k)\le D_{n,l,k}(c),
\label{eq:upperbound_Dlnk}
\end{equation}
where
\begin{eqnarray}
D_{n,l,k}(c)&:=&\sqrt{\frac{n(n+l-k)k}{(n+l)(n-k+c)(k-c)}}\nonumber\\
&&\times e^{\mu}2^{nh\left(\frac{k-c}{n}\right)-(n+l)h\left(\frac{k}{n+l}\right)+lh\left(\frac{c}{l}\right)},
\end{eqnarray}
\begin{equation}
\mu:=\frac1{6n}+\frac1{12}.
\label{eq:def_mu}
\end{equation}
\end{Lmm}

{\it Proof:}
By using the Stirling's formula
\begin{equation}
n!=\sqrt{2\pi n}\left(\frac{n}{e}\right)^ne^{\lambda_n}\ \ {\rm with}\ \ \frac1{12n+1}<\lambda_n<\frac1{12n},
\label{eq:stirlings_formula}
\end{equation}
we have
\begin{eqnarray}
\frac{{n \choose k-c}}{{n+l\choose k}}&=&\sqrt{\frac{n(n+l-k)k}{(n+l)(n-k+c)(k-c)}}
\label{eq:nchoosek-c_over_n+lchoosek}\\
&&\ \ \times e^{\mu'}\,2^{nh\left(\frac{k-c}{n}\right)-(n+l)h\left(\frac{k}{n+l}\right)}
\nonumber
\end{eqnarray}
where
\begin{eqnarray*}
\mu'&:=&\lambda_{n}-\lambda_{n-k+c}-\lambda_{k-c}-\lambda_{n+l}+\lambda_{n+l-k}+\lambda_{k}\\
&<&\lambda_{n}+\lambda_{n+l-k}+\lambda_{k}<\frac1{6n}+\frac1{12}
\end{eqnarray*}
for $\frac{1}{n+l}\le\frac{k}{n+l}\le\frac12$ and $l\le n$.
Combining (\ref{eq:nchoosek-c_over_n+lchoosek}) with $\sum_{i=0}^c{l \choose i}\le 2^{lh\left(\frac{c}{l}\right)}$ (see, e.g., Lemma 4.2.2 of \cite{Justesen}), we obtain (\ref{eq:upperbound_Dlnk}).
$\Box$

\begin{Lmm}
For $l\le n$, $c\le\bar{c}(k)$, and $\frac{k}{n+l}<\frac12$
\begin{equation}
nh\left(\frac{k-c}{n}\right)-(n+l)h\left(\frac{k}{n+l}\right)+lh\left(\frac{c}{l}\right)\le  -\frac1{2\ln2}\left(\frac{c-\bar{c}(k)}{\sigma(k)}\right)^2.
\label{eq:lmm_nh_n+lh_lh}
\end{equation}
\label{lmm:entropy_relation}
\end{Lmm}
{\it Proof:}
Since $h'''(x)$ decreases monotonically, we have
\begin{equation}
h(x)\le h(x_0)+h'(x_0)(x-x_0)+\frac12h''(x_0)(x-x_0)^2+\frac16h'''(x_0)(x-x_0)^3.
\label{eq:hx_hx0_derivatives}
\end{equation}
(Let $\tilde{h}(x)$ be the LHS minus the RHS.
It is easy to verify that $\tilde{h}(x_0)=\tilde{h}'(x_0)=\tilde{h}''(x_0)=\tilde{h}'''(x_0)=0$ and that $\tilde{h}'''(x)=h'''(x)-h'''(x_0)$ is a decreasing function.
Then by integrating $\tilde{h}'''(x)$ three times, one can show that $\tilde{h}(x)\le0$.)
Applying inequality (\ref{eq:hx_hx0_derivatives}) for $x_0=\frac{k}{n+l}$ and $x=\frac{k-c}{n}$, and also for $x=\frac{c}{l}$, we have
\begin{eqnarray}
\lefteqn{nh\left(\frac{k-c}{n}\right)-(n+l)h\left(\frac{k}{n+l}\right)+lh\left(\frac{c}{l}\right)}\nonumber\\
&\le&\frac12h''\left(\frac{k}{n+l}\right)\frac{n+l}{nl}(c-\bar{c}(k))^2\label{eq:ineq_nh_n+lh_lh}\\
&&\ +\frac16h'''\left(\frac{k}{n+l}\right)\left\{\frac1{n^{2}}-\frac1{l^{2}}\right\}(\bar{c}(k)-c)^3.\nonumber
\end{eqnarray}
Since $h'''\left(\frac{k}{n+l}\right)$, $\bar{c}(k)-c$, and $n-l$ are all non-negative by the conditions stated in the lemma, the second term on the right hand side is non-positive.
Then by noting 
\[
\frac{n+l}{nl}h''\left(\frac{k}{n+l}\right)
=-\frac1{(\ln2)\sigma(k)^2}\frac{n+l}{n+l-1}
\le-\frac1{(\ln2)\sigma(k)^2},
\]
we have Inequality (\ref{eq:lmm_nh_n+lh_lh}).
$\Box$

\begin{Lmm}
\label{lmm:C_upperbound}
If 
$c\le \bar{c}(k)$,
we have
\begin{equation}
\sqrt{\frac{n(n+l-k)k}{(n+l)(n-k+c)(k-c)}}
\le\sqrt{\frac{n+l}{n}}
\end{equation}
\end{Lmm}
{\it Proof:}
Let
\[
C(n,l,k,c):=\frac{n^2(n+l-k)k}{(n+l)^2(n-k+c)(k-c)}.
\]
Then it suffices to show $C\le1$ for $0\le c\le \bar{c}(k)$.

The function $f(k,c):=(n-k+c)(k-c)$ inside the square root is a concave parabola with its vertex at $c=k-\frac{n}{2}$.
This means that $f(k,c)\ge \min\left\{f(k,\bar{c}(k)), f(k,0)\right\}$, and thus $C(n,l,k,c)\le \max\{C(n,l,k,\bar{c}(k)),C(n,l,k,0)\}$.
Then it is straightforward to verity $C(n,l,k,\bar{c}(k))=1$ and $C(n,l,k,0)\le1$.
$\Box$

\begin{Lmm}
\label{lmm:upperbound_sum_Psi}
If $l\le n$, $1\le k$, $c\le \bar{c}(k)$ and $\frac{k}{n+l}\le\frac12$, we have
\begin{equation}
\sum_{i=0}^{c}P_{\rm hg}(i|k)
\le
e^{\mu}\sqrt{\frac{n+l}{n}}
\exp\left[-\frac12\left(\frac{c-\bar{c}(k)}{\sigma(k)}\right)^2\right].
\end{equation}
\end{Lmm}

{\it Proof:}
Combine Lemmas \ref{lmm:upperbound_Dlnk}, \ref{lmm:entropy_relation} and \ref{lmm:C_upperbound}.
$\Box$

\begin{Lmm}\label{Lmm:Chvatal}
If $0\le t$, $\bar{c}(k)-lt\le l/2$ and $\frac{k}{n+l}\le\frac12$,
\begin{equation}
\sum_{c=0}^{\bar{c}(k)-lt}P_{\rm hg}(c|k)
\le \exp\left[\frac{lt^2}{2}h''\left(\frac{k}{n+l}\right)\right].
\end{equation}
\end{Lmm}

{\it Proof:}
According to \cite{Chvatal},
\begin{eqnarray}
\sum_{i=0}^{\bar{c}(k)-lt}P_{\rm hg}(i|k)
&\le&
\left(\left(\frac{p}{p-t}\right)^{p-t}\left(\frac{1-p}{1-(p-t)}\right)^{1-(p-t)}\right)^l\nonumber\\
&=&2^{l[h(p-t)-h(p)+th'(p)]},
\end{eqnarray}
where $p=\frac{\bar{c}(k)}{l}=\frac{k}{n+l}$.
Since $h''(x)$ increases monotonically for $p-t\le x\le p\le1/2$, we have
\begin{eqnarray*}
h(p-t)\le h(p)+(-t)h'(p)+\frac{(-t)^2}2h''(p).
\end{eqnarray*}
That is,
\[
l[h(p-t)-h(p)+th'(p)]\le \frac{lt^2}2h''(p)
\]
$\Box$

\subsection{Upper and Lower Bounds on $\Phi(x)$}
\label{sec:upper_and_lowe_bounds_on_Phi}

\begin{Lmm}\label{lmm:normal dstr}
The normal distribution function, defined in (\ref{eq:def_normal_distribution_function}), is bounded as
\begin{equation}
\frac{\sqrt2}{\sqrt{x^2+2\pi}}e^{-x^2/2}\le \Phi(x)\le \frac{\sqrt2}{x}e^{-x^2/2}.
\end{equation}
\end{Lmm}

{\it Proof:}
According to Ref. \cite{Rus}, the function $\Phi(x)$ satisfies 
\begin{equation}
\tilde{g}_\pi(x)e^{-x^2/2}\le \Phi(x)\le \tilde{g}_4(x)e^{-x^2/2},
\label{eq:Ruskai_bound_Phi}
\end{equation}
where
\begin{equation}
\tilde{g}_k(x):=\frac{\sqrt2 k}{(k-1)x+\sqrt{x^2+2k}}.
\end{equation}
Then it is straightforward to show that for $k,x>0$,
\begin{equation}
\frac{\sqrt2}{\sqrt{x^2+2k}}\le \tilde{g}_k(x)\le \frac{\sqrt2}{x}.
\label{eq:bound_tilde_g}
\end{equation}
Combining (\ref{eq:Ruskai_bound_Phi}) and (\ref{eq:bound_tilde_g}), we obtain the lemma.
$\Box$

\begin{Lmm}\label{lmm:e-s2_le_epsilon}
If $\varepsilon=\Phi(s)$, and $2\le s$,
\begin{equation}
e^{-s^2}\le \frac\varepsilon2.
\end{equation}
\end{Lmm}

{\it Proof:}
From Lemma \ref{lmm:normal dstr},
\[
e^{-s^2}
\le e^{-s^2/2}\frac{\sqrt{s^2+2\pi}}{\sqrt2}\Phi(s)
= \sqrt{\frac{(s^2+2\pi)e^{-s^2}}{2}}\,\varepsilon.
\]
Then by noting $\frac{(s^2+2\pi)e^{-s^2}}{2}\le \frac14$ for $2\le s$, we obtain the lemma.
$\Box$

\subsection{Proof of Lemma \ref{lmm:upperbound_simple_bound}}
\label{Proof_lemma_simple_bound}

If $k/(n+l)\le \frac12$, by combining Lemmas \ref{lmm:upperbound_sum_Psi} and \ref{lmm:normal dstr}, we obtain
\[
\sum_{c=0}^{\lfloor\bar{c}-s\sigma\rfloor}P_{\rm hg}(i|k)
\le
\sqrt{\frac{n+l}{n}}
\sqrt{\frac{s^2+2\pi}{2}}e^{\mu}\varepsilon.
\]
On the other hand, if $k/(n+l)> \frac12$, by Lemma \ref{Lmm:Chvatal},
we have
\begin{eqnarray}
\sum_{c=0}^{c_{\max}}P_{\rm hg}(c|k)&\le& \sum_{c=0}^{c_{\max}}P_{\rm hg}\left(c|(n+l)/2\right)\nonumber\\
&\le &\exp\left[\frac{(1/2-0.12)^2l}2h''(1/2)\right]\nonumber\\
&\le& e^{-\frac{2}{5}l}\le e^{-\frac{s^2}{2}}.
\end{eqnarray}
Then by using Lemma \ref{lmm:normal dstr}, we have
\begin{equation}
\sum_{c=0}^{c_{\max}}P_{\rm hg}(i|k)< e^{-\frac12s^2}\le \sqrt{\frac{s^2+2\pi}{2}}\,\varepsilon.
\end{equation}

\section{Proof of Theorem 1}
\label{sec:proof_theorem_1}

\subsection{Proof of Case 1}
Since $p_{\rm sft}(k,c)=\frac{k-c}{n}\le \frac{k}{n}\le \frac{c_{\min}}{l}=p_{\rm smp}(c_{\min})$,
we have for arbitrary $c\in[0,l]$,
\begin{eqnarray*}
g(k,c)&=&nh\left(p_{\rm sft}(k,c)\right)-nh\left(\hat{p}_{{\rm sft},\varepsilon}\left(\max\{c+2,c_{\min}\}\right)\right)-D\\
&\le&nh\left(p_{\rm smp}(c_{\min})\right)-nh\left(\hat{p}_{{\rm sft},\varepsilon}(c_{\min})\right)-D.
\end{eqnarray*}
Further, from the concavity of $h(x)$ and from the monotonicity of $h'(x)$,
\begin{eqnarray*}
g(k,c)&\le&nh'(\hat{p}_{{\rm sft},\varepsilon}(c_{\min}))\left[p_{\rm smp}(c_{\min})-\hat{p}_{{\rm sft},\varepsilon}(c_{\min})\right]\\
&\le& nh'(\hat{p}_{{\rm sft},\varepsilon}(c_{\max}))\left[p_{\rm smp}(c_{\min})-\hat{p}_{{\rm sft},\varepsilon}(c_{\min})\right].
\end{eqnarray*}
Then by using Eq. (\ref{eq:def_hat_p_sft}) and by noting that $\left(p_{\rm smp}-\hat{p}_{\varepsilon}\right)^2 = 4\gamma \hat{p}_{\varepsilon}(1-\hat{p}_{\varepsilon})$ (see below Eq. (\ref{eq:def_gamma})),
\begin{eqnarray*}
g(k,c)&\le&-(n+l)h'(\hat{p}_{{\rm sft},\varepsilon}(c_{\max}))\left[\hat{p}_\varepsilon(c_{\min})-p_{\rm smp}(c_{\min})\right]-D\\
&=&-(n+l)h'(\hat{p}_{{\rm sft},\varepsilon}(c_{\max}))\\
&&\quad\times\sqrt{4\gamma}\sqrt{\hat{p}_\varepsilon(c_{\min})(1-\hat{p}_\varepsilon(c_{\min}))}-D\\
&=&-(1+4\gamma)s(\varepsilon)\beta\sigma\left((n+l)\hat{p}_\varepsilon(c_{\min})\right)-D\\
&\le&-\frac{\xi_{\min,\varepsilon} s(\varepsilon)^2}{\ln2}-D.
\end{eqnarray*}
The last inequality follows by noting that $nc_{\min}/l\le (n+l)\hat{p}_\varepsilon(c_{\min})\le(n+l)\hat{p}_\varepsilon(c_{\max})$, and thus $-\sigma\left((n+l)\hat{p}_\varepsilon(c_{\max})\right)\le -\sigma\left(nc_{\min}/l\right)$.
Then by using Lemma \ref{lmm:e-s2_le_epsilon}, we have for $1<\xi_{\min,\varepsilon}$ and $D=1$,
\[
S_{\rm pa}(k,c)\le2^{[g(k,c)]^-+1}\le 2e^{-\xi_{\min,\varepsilon} s(\varepsilon)^2}< \varepsilon\\
\]
$\Box$

\subsection{Proof of Case 2}
This part is immediate from the following lemma.
\begin{Lmm}
\label{lmm:g_upperbound}
Suppose $1\le l\le n$, $4\gamma\le1$.
Then, for any integer $k$, any real number $\varepsilon>0$ and any $c\in[\,\bar{c}(k)-s(\varepsilon)\sigma(k),\,c_{\max}\,]$, we have 
\begin{equation}
g(k,c)\le -\beta\left(c-(\bar{c}(k)-s(\varepsilon)\sigma(k))+1\right)-D,
\end{equation}
with $\beta$ defined in (\ref{eq:def_beta}).
\end{Lmm}

{\it Proof:}
With $h(x)$ being concave, and with $\hat{p}_{{\rm sft},\varepsilon}(c)$ increasing monotonically,
\begin{eqnarray}
g(k,c)&\le&-nh'(\hat{p}_{{\rm sft},\varepsilon}(c+2))\left(\hat{p}_{{\rm sft},\varepsilon}(c+2)-p_{\rm sft}(k,c)\right)-D\nonumber\\
&\le& -nh'(\hat{p}_{{\rm sft},\varepsilon}(c_{\max}+2))\left(\hat{p}_{{\rm sft},\varepsilon}(c+2)-p_{\rm sft}(k,c)\right)-D\nonumber.
\end{eqnarray}
The quantity $\hat{p}_{{\rm sft},\varepsilon}(c+2)-p_{\rm sft}(k,c)$ on the right hand side can be bounded as follows.
First note $\hat{p}_{{\rm sft},\varepsilon}(\bar{c}-s\sigma)-p_{\rm sft}(k,\bar{c}-s\sigma)=0$ by the definition of $\hat{p}_{{\rm sft},\varepsilon}(c)$, given in (\ref{eq:def_hat_p}) and (\ref{eq:def_hat_p_sft}).
Also by the definition of $\hat{p}_{{\rm sft},\varepsilon}(c)$, we have that $\frac{d\hat{p}_{{\rm sft},\varepsilon}}{dc}\ge\frac1{1+4\gamma}\frac{n+l}{nl}-\frac1n$,
and that $\frac{\partial p_{\rm sft}}{\partial c}=-\frac1n$ by the definition of $p_{\rm sft}(k,c)$;
hence $\frac{\partial}{\partial c}(\hat{p}_{{\rm sft},\varepsilon}-p_{\rm sft})\ge \frac1{1+4\gamma}\frac{n+l}{nl}$.
Thus $\hat{p}_{{\rm sft},\varepsilon}(\bar{c}-s\sigma+2)-p_{\rm sft}(k,\bar{c}-s\sigma+2)\ge \frac2{1+4\gamma}\frac{n+l}{nl}$.
Then for $\bar{c}(k)-s(\varepsilon)\sigma(k)\le c$, we have
\begin{eqnarray}
\lefteqn{\hat{p}_{{\rm sft},\varepsilon}(c+2)-p_{\rm sft}(k,c)}\\
&=&(\hat{p}_{{\rm sft},\varepsilon}(c+2)-p_{\rm sft}(k,c+2))+(p_{\rm sft}(k,c+2)-p_{\rm sft}(k,c))\\
&\ge&\frac1{1+4\gamma}\frac{n+l}{nl}\left(c-(\bar{c}-s\sigma)+2\right)-\frac2{n}\\
&\ge&\frac1{1+4\gamma}\frac{n+l}{nl}\left(c-(\bar{c}-s\sigma)+1\right).
\end{eqnarray}
$\Box$

Plugging this upper bound on $g(k,c)$ (for $D=1$) to $S_{\rm pa}(k,c)$ (given in (\ref{eq:upper_bound_S_pa}) and (\ref{eq:def_g_k_c})), we obtain Case 2 of Theorem 1.

\section{Proof of Theorem \ref{thm:exact_upperbound}}
\label{sec:proof_theorem_3}
Next we prove Theorem \ref{thm:exact_upperbound} starting from Theorem 1.
In the following, $s(\varepsilon)$ is simplified to $s$.

Under the conditions of Case 1 of Theorem 1, inequality (\ref{eq:case1_upperbound}) holds independently of the normal approximation, and thus we readily see that (\ref{eq:Theorem_3_exact_upperbound}) holds.

\begin{Lmm}\label{lmm:upperbound_P_k_c_1}
If $1\le l\le n$, $1\le k$, $c\le \bar{c}(k)$ and $\frac{k}{n+l}\le1/2$, we have
\begin{equation}
P_{\rm hg}(c|k)\le \frac{e^{\mu+\nu}}{\sqrt{2\pi}\sigma((n+l)c/l)}
\exp\left[-\frac12\left(\frac{c-\bar{c}(k)}{\sigma(k)}\right)^2\right],
\end{equation}
with $\mu$ defined in (\ref{eq:def_mu}), and
\begin{equation}
\nu:= \frac{1}{12l}+\frac1{2(n+l-1)}.
\label{eq:def_nu}
\end{equation}
\end{Lmm}

{\it Proof:}
By using the Stirling's formula (\ref{eq:stirlings_formula}), we have
\begin{equation}
{l\choose c}\le\sqrt\frac{n}{n+l-1}
\frac1{\sqrt{2\pi}\sigma((n+l)c/l)}e^{\nu'}2^{lh(c/l)},
\label{eq:l_choose_c_upperbound}
\end{equation}
where
\begin{equation}
\nu'=\lambda_l-\lambda_{l-c}-\lambda_c\le\lambda_l<\frac1{12l}.
\end{equation}
Then by combining Inequality (\ref{eq:l_choose_c_upperbound}) with
(\ref{eq:nchoosek-c_over_n+lchoosek}) and (\ref{eq:lmm_nh_n+lh_lh}), and by using Lemma \ref{lmm:C_upperbound},
we obtain
\[
P_{\rm hg}(c|k)\le \frac{e^{\mu+\frac{1}{12l}}}{\sqrt{2\pi}\sigma((n+l)c/l)}\sqrt{1+\frac{1}{n+l-1}}
\exp\left[-\frac12\left(\frac{c-\bar{c}(k)}{\sigma(k)}\right)^2\right].
\]
Then by noting
\[
\sqrt{1+\frac{1}{n+l-1}}
\le
\sqrt{\exp\left(\frac{1}{n+l-1}\right)}
=\exp\left(\frac{1}{2(n+l-1)}\right),
\]
we obtain the lemma.
$\Box$

\begin{Lmm}
\label{lmm:upperbound_P_k_c_2}
If $l\le n$, $1\le c_{\min}$, $nc_{\min}/l\le k$, $\bar{c}(k)-s\sigma(k)\le c\le \bar{c}(k)$ and $\frac{k}{n+l}\le1/2$, we have
\begin{equation}
P_{\rm hg}(c|k)\le \frac{e^{\mu+\nu}}{\sqrt{1-\frac{s}{\sqrt{c_{\min}}}}}
\frac{1}{\sqrt{2\pi}\sigma(k)}
\exp\left[-\frac12\left(\frac{c-\bar{c}(k)}{\sigma(k)}\right)^2\right],
\end{equation}
with $\mu$, $\nu$ defined in (\ref{eq:def_mu}), (\ref{eq:def_nu})
\end{Lmm}

{\it Proof:}
From the definition of $\sigma(k)$, we have
\[
\frac{\sigma(k)}{\sigma(k(1-s\sigma(k)/\bar{c}(k)))}
\le \frac{1}{\sqrt{1-s\sigma(k)/\bar{c}(k)}}.
\]
By noting that $nc_{\min}/l\le k$, we have
\begin{eqnarray*}
\frac{\sigma(k)}{\bar{c}(k)}&=&\sqrt{\frac{n}{l(n+l-1)}\left(\frac{n+l}{k}-1\right)}\\
&\le&\sqrt{\frac{n}{l(n+l-1)}\left(\frac{l(n+l)}{nc_{\min}}-1\right)}\\
&=&\sqrt{\frac{n}{l(n+l-1)}\left(\frac{l(n+l)-nc_{\min}}{nc_{\min}}\right)}\\
&\le&\sqrt{\frac{n}{l(n+l-1)}\left(\frac{l(n+l-1)}{nc_{\min}}\right)}\\
&\le&\frac1{\sqrt{c_{\min}}}.
\end{eqnarray*}
Hence $1-\frac{s\sigma(k)}{\bar{c}(k)}\ge 1-\frac{s}{\sqrt{c_{\min}}}$.
The assumption yields that
$(n+l)c/l \ge k(1-s\sigma(k)/\bar{c}(k))$,
which implies
\[
\frac{\sigma(k)}{\sigma((n+l)c/l)}
\le
\frac{\sigma(k)}{\sigma(k(1-s\sigma(k)/\bar{c}(k)))}
\le \frac{1}{\sqrt{1-\frac{s}{\sqrt{c_{\min}}}}}.
\]
Combining this inequality with Lemma \ref{lmm:upperbound_P_k_c_1}, we obtain Lemma \ref{lmm:upperbound_P_k_c_2}.
$\Box$

\subsection{Proof of Case 2}

If $\frac{k}{n+l}\ge\frac12$, this case can be proved by exactly the same argument as in \ref{Proof_lemma_simple_bound} 
(Note here that the condition $s^2\le c_{\min}\le c_{\max}\le 0.12l$, appearing in Theorem 3, implies $\frac54s^2\le l$).
Hence in this subsection, we assume that $\frac{k}{n+l}<\frac12$.
We also assume that $1\le k$, because the case $k=0$ is already considered in Case 1 of Theorem 1.

First we divide the right hand side of (\ref{eq:case2_upperbound}) into three parts,
\begin{eqnarray}
\lefteqn{\sum_{c=0}^{c_{\max}}P_{\rm hg}(c|k)S_{\rm pa}(k,c)}\nonumber\\
&\le&\sum_{c=0}^{\lfloor\bar{c}(k)-s\sigma(k)\rfloor}P_{\rm hg}(c|k)
+\sum_{c=\lfloor\bar{c}(k)-s\sigma(k)\rfloor+1}^{\lfloor\bar{c}(k)\rfloor-1} P_{\rm hg}(c|k)S_{\rm pa}(k,c)\nonumber\\
&&+\sum_{c=\lfloor\bar{c}(k)\rfloor}^{c_{\max}} P_{\rm hg}(c|k)S_{\rm pa}(k,c)
\label{eq:exact_case2_divided}.
\end{eqnarray}
The first term on the right hand side can be bounded from above by Lemma 1.
The second term can be bounded as
\begin{eqnarray*}
\lefteqn{\sum_{c=\lfloor\bar{c}(k)-s\sigma(k)\rfloor+1}^{\lfloor\bar{c}(k)\rfloor-1} P_{\rm hg}(c|k)S_{\rm pa}(k,c)}\\
&\le&\sum_{c=\lfloor\bar{c}(k)-s\sigma(k)\rfloor+1}^{\lfloor\bar{c}(k)\rfloor-1} P_{\rm hg}(c|k)2^{-\beta(c-(\bar{c}(k)-s\sigma(k))+1)}\\
&\le&\frac{e^{\mu+\nu}}{\sqrt{1-\frac{s}{\sqrt{c_{\min}}}}}\frac{1}{\sqrt{2\pi}\sigma(k)}\\
&&\times\sum_{c=\lfloor\bar{c}(k)-s\sigma(k)\rfloor+1}^{\lfloor\bar{c}(k)\rfloor-1}
\exp\left[-\frac12\left(\frac{c-\bar{c}(k)}{\sigma(k)}\right)^2\right]\\
&&\hspace{15ex}\times 2^{-\beta(c-(\bar{c}(k)-s\sigma(k))+1)}\\
&\le&
\frac{e^{\mu+\nu}}{\sqrt{1-\frac{s}{\sqrt{c_{\min}}}}}
\frac{1}{\sqrt{2\pi}}
\int_{-s}^{\infty}dx\ e^{-x^2/2}2^{-\beta\sigma(k)(x+s)}\\
&\le&
\frac{e^{\mu+\nu}}{\sqrt{1-\frac{s}{\sqrt{c_{\min}}}}}
I_2\left(\xi_{\varepsilon}(k)\right).
\end{eqnarray*}
Then $I_2\left(\xi_{\varepsilon}(k)\right)$ appearing in the last line can be bounded by Inequality (\ref{eq:upper_bound_I_2}).
(Note that the argument in the paragraph of Inequality (\ref{eq:upper_bound_I_2}) does not rely on the normal approximation.)

The third summation on the right hand side of (\ref{eq:exact_case2_divided}) can be bounded as
\begin{eqnarray*}
\lefteqn{\sum_{c=\lfloor\bar{c}(k)\rfloor}^{c_{\max}+1}
P_{\rm hg}(c|k)2^{-\beta(c-(\bar{c}(k)-s\sigma(k))+1)}}\\
&\le&
\sum_{c=\lfloor\bar{c}(k)\rfloor}^{ c_{\max}+1} P_{\rm hg}(c|k)2^{-\beta(c-(\bar{c}(k)-s\sigma(k))+1)}
\\
&\le&2^{-\beta\sigma(k) s}\le e^{-\xi_{\varepsilon}(k) s^2}\le \varepsilon^{\xi_{\varepsilon}(k)}\le\varepsilon^2.
\end{eqnarray*}




\section{Proof of Theorem \ref{thm:gauss_asymptotic}:}\label{proof-gauss_asymptotic}
First, we fix arbitrary $\varepsilon' > \varepsilon$.
Since the function $h(x)$ and its derivative $h'(x)$ are uniformly continuous in the range $[p_{\min},p_{\max}]$,
there exists an integer $N$ such that
$
\lceil nh\left(\hat{p}_{{\rm sft},\varepsilon'}(c+1)\right)\rceil+1
\le
\lceil nh \left(p_{\rm smp}(c)\right) + \sqrt{n} h' \left(p_{\rm smp}(c)\right)
\sqrt{\frac{p_{\rm smp}(c)(1-p_{\rm smp}(c)) (1+t)}{4t}}
s(\varepsilon) \rceil
$ for $n \ge N$ and $l \ge tn$.
Using Theorem 1 of \cite{Lahiri},
we can choose constants $C_1$ and $C_2$ such that
$P_{\rm hg}(c|k) \le  
\frac1{\sqrt{2\pi}}\int_{\zeta_c}^{\zeta_{c+1}}e^{-x/2}dx
+ \frac{C_1 (1+ \zeta_c^2)}{\sigma_{n,l}(k)} \exp(-C_2 \zeta_c^2)$.
Here note that the constants $C_1$ and $C_2$ are different from those defined in Theorem 1 of \cite{Lahiri}.

Using $C_3:=\int_{-\infty}^\infty
C_1 (1+ x^2)\exp(-C_2 x^2)$,
we obtain
\begin{eqnarray}
\sum_{c=0}^{np_{\max} } \frac{C_1 (1+ \zeta_c^2)}{\sigma_{n,l}(k)} \exp(-C_2 \zeta_c^2)
\min \left\{2^{-\beta\left(c-(\bar{c}-s(\varepsilon)\sigma)\right)},1\right\}
\le \frac{C_3 }{\sigma_{n,l}(k)}.
\end{eqnarray}
Hence, Theorem \ref{thm:gauss_upperbound} yields that
\begin{eqnarray}
P_{{\rm ph},n,l}\le
(1+\delta'_n)\varepsilon' 
+ \frac{C_3}{\min_{k: np_{\min} \le k \le (n+l) \left(\hat{p}_{{\rm sft},\varepsilon'}(l p_{\max}+1)\right)}\sigma_{n,l}(k)}.
\end{eqnarray}
where $\delta'_n$
is the maxumum of $\delta$ given in Theorem \ref{thm:gauss_upperbound} with the condition $l \ge tn$.

Since 
$\min_{l:l \ge tn }
\min_{k: np_{\min} \le k \le (n+l) \left(\hat{p}_{{\rm sft},\varepsilon'}(l p_{\max}+1)\right)}\sigma_{n,l}(k)
\to \infty$ as $n \to \infty$,
we obtain
$\lim_{n \to \infty} \max_{l:l \ge tn }
\frac{C_3}{\min_{k: np_{\min} \le k \le (n+l) \left(\hat{p}_{{\rm sft},\varepsilon'}(l p_{\max}+1)\right)}\sigma_{n,l}(k)}
=0$.
Also we can show that $\delta'_n \to 0$.
Thus, we obtain 
$\lim_{n \to \infty} \max_{l:l \ge tn } P_{{\rm ph},n,l} \le \varepsilon'$.
Since $\varepsilon'$ is an arbitrary real number satisfying that $\varepsilon'> \varepsilon$.
Hence,
$\lim_{n \to \infty} \max_{l:l \ge tn } P_{{\rm ph},n,l} \le \varepsilon$.
$\Box$

\end{document}